\documentclass[longauth]{aa}  

\usepackage{graphicx}
\usepackage{txfonts}
\usepackage{lscape}
\usepackage{natbib}
\bibpunct{(}{)}{;}{a}{}{,} 
\usepackage{hyperref}
\hypersetup{colorlinks,linkcolor=red,citecolor=blue,urlcolor=blue}

\def\teff{$T_{\mathrm{eff}}$}  
\def\logg{$\log g$}   
\def\feh{[Fe/H]}     
\def\micro{$\xi$}    

\hypersetup{draft}

\begin{document} 

   \title{The {\it Gaia}-ESO Survey: Spectroscopic-asteroseismic analysis of K2 stars in {\it Gaia}-ESO}

  \subtitle{The K2 Galactic Caps Project}

   \author{C.~C.~Worley\inst{1}
          \and
          P. Jofr\'e\inst{2}
          \and 
          B. Rendle\inst{3}
          \and 
          A. Miglio\inst{3}
          \and
          L. Magrini\inst{4}
          \and
          D. Feuillet\inst{5,}\inst{12}
          \and
          A. Gavel\inst{6}
          \and
          R. Smiljanic\inst{7}
          \and
          K. Lind\inst{5,}\inst{6,}\inst{21}
          \and
          A. Korn\inst{6}
          \and
          G. Gilmore\inst{1}
          \and
          S. Randich\inst{4}
          \and
          A. Hourihane\inst{1}
          \and
          A. Gonneau\inst{1}
          \and
          P. Francois\inst{8}
          \and
          J. Lewis\inst{1}
          \and
          G. Sacco\inst{4}
          \and
          A. Bragaglia\inst{11}
          \and
          U. Heiter\inst{6}
          \and
          S. Feltzing\inst{12}
          \and
          T. Bensby\inst{12}
          \and
          M. Irwin\inst{1}
          \and
          E. Gonzalez Solares\inst{1}
          \and
          D. Murphy\inst{1}
          \and
          A. Bayo\inst{19,}\inst{22}
          \and
          L. Sbordone\inst{17}
          \and
          T. Zwitter\inst{18}
          \and
          A.~C. Lanzafame\inst{10}        
          \and
          N. Walton\inst{1}
          \and
          S. Zaggia\inst{9}
          \and
          E.~J. Alfaro\inst{13}
          \and
          L. Morbidelli\inst{4}
          \and
          S. Sousa\inst{14}
          \and
          L. Monaco\inst{15}
          \and
          G. Carraro\inst{16}
          \and
          C. Lardo\inst{20}
          }
          
   \institute{Institute of Astronomy, University of Cambridge, Madingley Road, Cambridge, CB3 0HA, UK, \email{ccworley@ast.cam.ac.uk}
         \and
            N\'ucleo de Astronom\'{i}a, Universidad Diego Portales, Av. Ej\'{e}rcito 441, Santiago de Chile,
             \email{paula.jofre@mail.udp.cl}
           \and 
           School of Physics and Astronomy, University of Birmingham, Birmingham B15 2TT, UK
           \and 
           INAF - Osservatorio Astrofisico di Arcetri, Largo E. Fermi 5, 50125, Florence, Italy
           \and 
           Max-Planck Institut f\"{u}r Astronomie, K\"{o}nigstuhl 17, 69117 Heidelberg, Germany
           \and 
           Observational Astrophysics, Division of Astronomy and Space Physics, Department of Physics and Astronomy, Uppsala University, Box 516, SE-751 20 Uppsala, Sweden
           \and 
           Nicolaus Copernicus Astronomical Center, Polish Academy of Sciences, ul. Bartycka 18, 00-716, Warsaw, Poland
           \and
           GEPI, Observatoire de Paris, CNRS, Universit\'{e} Paris Diderot, 5 Place Jules Janssen, 92190 Meudon, France
           \and
           INAF - Padova Observatory, Vicolo dell'Osservatorio 5, 35122 Padova, Italy
           \and
           Dipartimento di Fisica e Astronomia, Sezione Astrofisica, Universit\'{a} di Catania, via S. Sofia 78, 95123, Catania, Italy
           \and
           INAF - Osservatorio di Astrofisica e Scienza dello Spazio di Bologna, via Gobetti 93/3, 40129, Bologna, Italy
           \and
           Lund Observatory, Department of Astronomy and Theoretical Physics, Box 43, SE-221 00 Lund, Sweden
           \and
           Instituto de Astrof\'{i}sica de Andaluc\'{i}a-CSIC, Apdo. 3004, 18080, Granada, Spain
           \and
           Instituto de Astrof\'isica e Ci\^encias do Espa\c{c}o, Universidade do Porto, CAUP, Rua das Estrelas, 4150-762 Porto, Portugal
           \and
           Departamento de Ciencias Fisicas, Universidad Andres Bello, Fernandez Concha 700, Las Condes, Santiago, Chile
           \and
           Dipartimento di Fisica e Astronomia, Universit\`a di Padova, Vicolo dell'Osservatorio 3, 35122 Padova, Italy
           \and
           European Southern Observatory, Alonso de Cordova 3107 Vitacura, Santiago de Chile, Chile
           \and
           Faculty of Mathematics and Physics, University of Ljubljana, Jadranska 19, 1000, Ljubljana, Slovenia
           \and
           Instituto  de  F\'isica  y Astronom\'ia,  Facultad  de Ciencias,  Universidad de Valpara\'iso, Av. Gran Breta\~na 1111, 5030 Casilla, Valpara\'iso, Chile
           \and
           Laboratoire d'astrophysique, Ecole Polytechnique F\'ed\'erale de Lausanne (EPFL), Observatoire de Sauverny, CH-1290 Versoix, Switzerland
           \and
           Department of Astronomy, Stockholm University, AlbaNova, Roslagstullbacken 21, SE-10691 Stockholm, Sweden
             \and
             N\'ucleo Milenio de Formaci\'on Planetaria - NPF, Universidad de Valpara\'iso, Av. Gran Breta\~na 1111, Valpara\'iso, Chile
                          }
    \date{Received ; accepted }


  \abstract
   {The extensive stellar spectroscopic datasets that are available for studies in Galactic Archeaology thanks to, for example, the {\it Gaia}-ESO Survey, now benefit from having a significant number of targets that overlap with asteroseismology projects such as {\it Kepler}, K2, and CoRoT. Combining the measurements from spectroscopy and asteroseismology allows us to attain greater accuracy with regard to the stellar parameters needed to  characterise the stellar populations of the Milky Way.}
   {The aim of this {\it Gaia}-ESO Survey special project is to produce a catalogue of self-consistent stellar parameters by combining measurements from high-resolution spectroscopy and precision asteroseismology.}
   {We carried out an iterative analysis of 90 K2@{\it Gaia}-ESO red giants. The spectroscopic values of \teff{} were used as input in the seismic analysis to obtain \logg{} values. The seismic estimates of \logg{} were then used to re-determine the spectroscopic values of \teff{} and \feh{}. Only one iteration was required to obtain parameters that are in good agreement for both methods and, thus, to obtain the final stellar parameters. A detailed analysis of outliers was carried out to ensure a robust determination of the parameters. The results were then combined with {\it Gaia} DR2 data to compare the seismic \logg{} with a parallax-based \logg{} and to investigate instances of variations in the velocity and possible binaries within the dataset.}
   {This analysis produced a high-quality catalogue of stellar parameters for 90 red giant stars from K2@{\it Gaia}-ESO that were determined through iterations between spectroscopy and asteroseismology. We compared the seismic gravities with those based on Gaia parallaxes to find an offset which is similar to other studies that have used asteroseismology. Our catalogue also includes spectroscopic chemical abundances and radial velocities, as well as indicators for possible binary detections.}
   {%
   }


   \maketitle
\section{Introduction}
The characterisation of the Milky Way stellar populations in studies of Galactic Archaeology has been greatly advanced with the recently acquired wealth of high-quality spectroscopic data that are now available for hundreds of thousands of stars in our Galaxy.
The recent Data Release 2 for the European Space Agency (ESA) space mission, Gaia, has led to the publication of highly accurate astrometry for over a billion stars in the Milky Way \citep{GaiaDR2}.

In anticipation of this astrometric mecca, a suite of stellar spectroscopic surveys of high-resolution, such as {\it Gaia}-ESO \citep{gaiaesomess}, APOGEE \citep{apogee}, and GALAH \citep{2015MNRAS.449.2604D} were created, initiating a new era of large databases containing the spectra and scientific measurements for hundreds of thousands of stars. The next wave of surveys, such as WEAVE \citep{Dalton2012}, 4MOST \citep{2019Msngr.175....3D}, MOONS \citep{MOONS2016}, and MSE \citep{Szeto2018MSE} will expand the coverage of these databases into the millions.

However, the robustness of the stellar parameters determined for these large spectroscopic datasets depends, in particular, on the accuracy of the parameters of just a few small samples of reference stars \citep[see discussion in e.g.][]{Jofre18}. A stellar reference set commonly used for the validation and verification of automated stellar parameterisation pipelines is the {\it Gaia} FGK Benchmark Stars \citep{jofre14, heiter15}. These are very bright stars but they number only 36 \citep[][for the latest list]{jofre18gbs} and so, they sparsely sample the FGK stellar parameter space. Reference sets are also drawn from stellar spectral libraries, in particular the ELODIE library \citep{elodie}, compilations of high quality (but inhomogeneous) literature values such as PASTEL \citep{pastel}, and stars that are well-known members of open clusters or globular clusters. 

These relatively small reference sets are being used to define the parameter scale zero-point upon which the large scale analyses are then based \citep[e.g.,][]{pancino17, kunder17}. They are therefore crucial for determining the accurate absolute values of the reported stellar parameters (effective temperature \teff{}, surface gravity \logg{}, metallicity \feh{}, microturbulence $\xi$) for these large-scale surveys. They are also essential for effective comparison of the survey datasets with stellar and galactic evolution models as this is key to making a straightforward comparison and combination of data from multiple surveys analysed by different pipelines. In the future, parameters for these reference sets derived within {\it Gaia}-ESO and other surveys can be compared to results drawn from more sophisticated models of stellar atmospheres that may include advances in non-LTE, 3D, and dynamical atmospheres, for example.

However, the relatively small number of reference stars stands as a problem given the required stellar parameter space is not well-sampled for a comprehensive analysis by the automated pipelines. Greater coverage is needed all the way from cool pre-main sequence stars to hot OB stars. New stellar reference sets of independently determined stellar parameters are required to keep up with the demands of upcoming large scale surveys. The work presented here seeks to define a new sample of reference stars generated by combining the stellar parameters of \teff{} and \feh{} from spectroscopy with \logg{} from asteroseismology.

Today, the availability of datasets with a wide sky coverage has resulted in many overlapping targets between research fields and presents, thus, an opportunity to make simultaneous use of the collective strengths of multiple types of analyses. This is the case here, among the thousands of targets with high-resolution and high-quality stellar spectra, where there is also asteroseismic information for some thanks to the dedicated monitoring of their oscillations by {\it Kepler} \citep{2010Sci...327..977B}, CoRoT \citep{2006ESASP1306...33B} and, recently, by K2 \citep{2014PASP..126..398H}. These asteroseismic measurements of the interior of stars combined with the spectroscopic `exterior' measurements of the same stars and the direct measurement of the parallax of each star, are part of the unfolding revolution in Galactic Archaeology \citep{2017AN....338..644M}. To take advantage of and further develop this multi-analysis approach, the {\it Gaia}-ESO Survey observed several hundreds of stars in the K2 Campaign 3 (C3) field, located towards the South Galactic pole. 

The {\it Gaia}-ESO Survey is an ESO Large Public Spectroscopic survey designed to target over 100,000 in the key stellar populations of the Milky Way \citep{gaiaesomess}. It has made use of a large range of analysis methods and thus developed key homogenisation procedures to bring all the results together in a robust single star catalogue, the final data release of which is currently underway. {\it Gaia}-ESO observed stars using the medium-resolution spectrograph GIRAFFE (R$\sim$20,000) and the high-resolution spectrograph UVES (R$\sim$47,000) on the VLT. 

K2 is a re-purposing of the {\it Kepler} satellite for which the science goals are focussed on the detection of the variations in the light curves of stars in 19 fields along the ecliptic to look for transiting exo-planets. At the time this special project began, 90 giant stars with oscillations detected by K2 were identified within the sample of {\it Gaia}-ESO stars observed in C3 with medium- and high- resolution spectroscopy. This sample is referred to hereafter as K2@{\it Gaia}-ESO.

In this paper, which is part of the series of the K2 Galactic Caps Project \citep[see also][hereafter Paper I]{rendlek2ges} and is, in particular, a collaboration with {\it Gaia}-ESO, we describe the process of obtaining accurate atmospheric parameters that are consistent with the results of asteroseismology. We use our results to explore age dependencies with abundance ratios and examine potential binary stars in our sample. This sample will further provide a good opportunity for the study of any possible offsets between standard spectroscopic parameters and future sets of parameters derived from more sophisticated models of stellar atmospheres (e.g. non-LTE, 3D, dynamical atmospheres). 

In Sect.~\ref{sect:data}, we describe in more detail the data we use for this work. In Sect.~\ref{sect:parameters}, we describe the process we used to determine atmospheric parameters iteratively between spectroscopy and asteroseismology. In Sect.~\ref{sect:abundances}, we present our chemical abundance results. The comparison to \logg{} based on {\it Gaia} parallaxes is presented in Sect.~\ref{sect:parallaxes}. Our findings regarding binary stars are described in Sect.~\ref{sect:binaries} and the final discussion and conclusions are presented in Sect.~\ref{sect:conclusion}.

\section{The {\it Gaia}-ESO Survey sample of K2 stars}
\label{sect:data}

Targets within the K2 C3 field were prepared as part of the {\it Gaia}-ESO observing programme. This resulted in 496 observations that were available for analysis. The initial set of targets were observed in May and June 2016 and, thus,  added into the internal Data Release (iDR) 5 data analysis cycle of {\it Gaia}-ESO that had begun at the start of May 2016. The rest of the fields were observed in October 2016 and are part of iDR6.

In total there were 231 targets observed using UVES 580 (blue and red arms) and 265 targets observed using the HR10 and HR21 setups of GIRAFFE. Of these, 182 UVES targets and 133 GIRAFFE targets had been included as part of the iDR5 analysis. 

\begin{figure}[t]
\centering
\includegraphics[width=9.8cm]{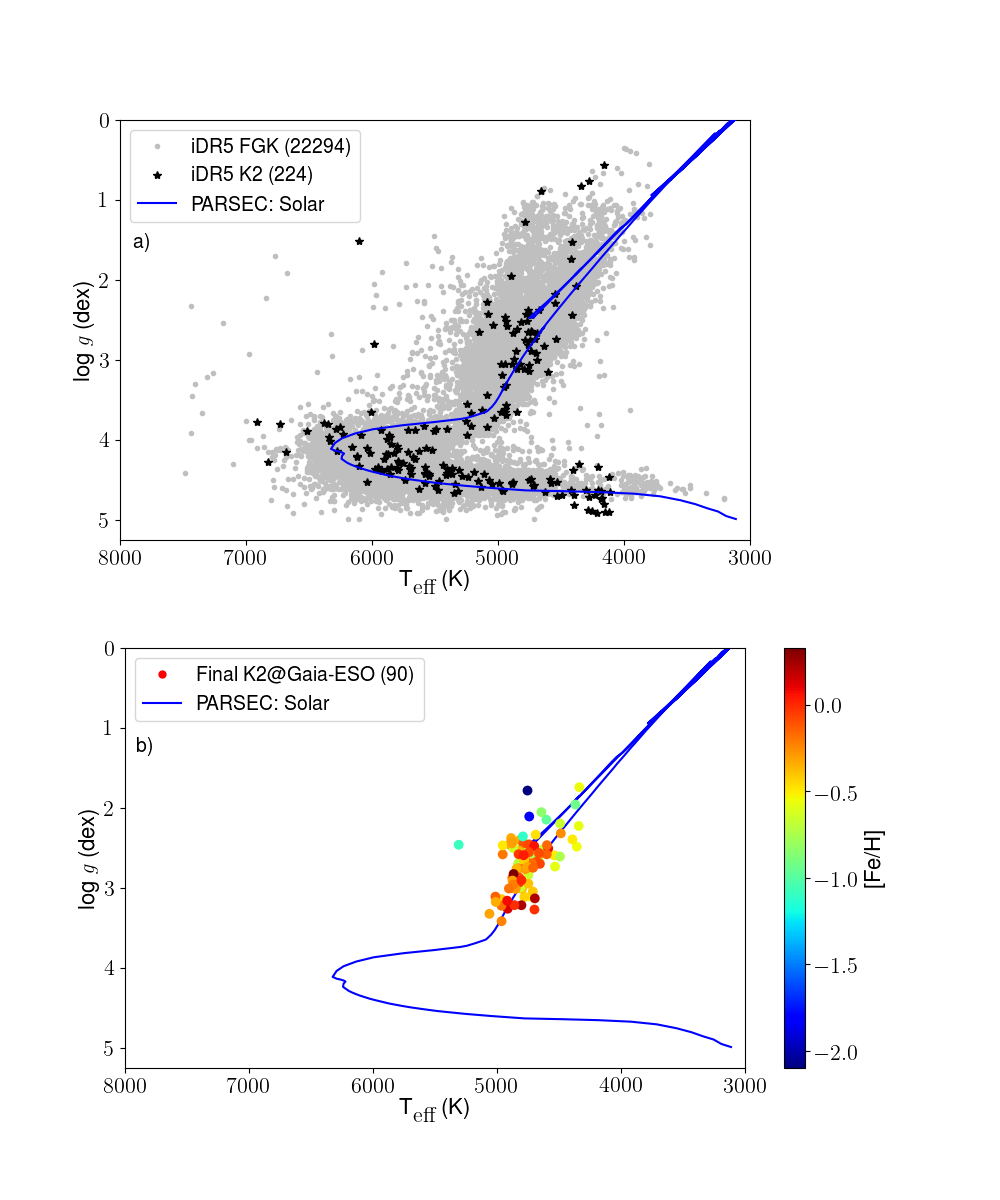}
\caption{a) HR Diagram of iDR5 FGK stars (grey points), K2 C3 stars analysed in {\it Gaia}-ESO iDR5 (black stars) and PARSEC Solar track. b) Kiel diagram of the final set of K2@{\it Gaia}-ESO stars for which we obtained spectroscopic parameters as analysed in this paper, with PARSEC Solar track.}
\label{fig:initialHRD}
\end{figure}

\subsection{Cross-match to K2}
The K2 C3 stars that are part of the {\it Gaia}-ESO survey were observed by K2 after they were observed by {\it Gaia}-ESO. Therefore, it was not known at the time when {\it Gaia}-ESO was observing them how many would ultimately have asteroseismic detections. At the start of this spectroscopic analysis, 90 of the 496 targets were identified as having K2 asteroseismic detections; 28 of these were observed with GIRAFFE, 62 were observed with UVES.  It is possible that more of the full sample will have asteroseismic detections as the K2 analysis advances, but we leave this consideration to future works.

Figure~\ref{fig:initialHRD}a shows the HR diagram of the {\it Gaia}-ESO iDR5 FGK stars (S/N$>$30) and the 224 K2 targets (out of 496) observed by {\it Gaia}-ESO that were analysed in iDR5 with this signal-to-noise ratio (S/N) cut. A PARSEC stellar track of Solar metallicity and age is also shown \citep{PARSEC}. Figure~\ref{fig:initialHRD}b shows the final stellar parameters of the 90 K2@{\it Gaia}-ESO sample analysed in this work with the [Fe/H] colourmap. Details of the seismic analysis for these targets can be found in Paper~I.

\subsection{Preliminary spectroscopic parameters}
It was important to initiate the iterations between the spectroscopic and seismic analyses from the best starting point possible in \teff{}. As not all of the K2@{\it Gaia}-ESO targets were observed in time to be included in iDR5, it was necessary to compile the rest of the {\it preliminary} stellar parameter set from a variety of other sources including: a photometric \teff{}; parameters associated with the synthetic template used in the radial velocity determination in the reduction pipelines; and parameters derived using an available {\it Gaia}-ESO node analysis. These values are provided for just the 90 K2@{\it Gaia}-ESO stars with asteroseismic detections in Table~\ref{tab:ges_k2_prelim} along with the {\it Gaia}-ESO CNAME, EPIC identifier and the instrument with which the spectrum was observed for {\it Gaia}-ESO. 


For both samples, the infrared flux method (IRFM) calibrations of \cite{Ramirez2005} were used to estimate a photometric \teff{} using the APASS V magnitude and the 2MASS K$_s$ magnitude. The \feh{} from the iDR5 recommended parameters were used as input to the IRFM calibration equations where possible. Otherwise the \feh{} from the GIRAFFE radial velocity determination \citep{gilmoreges} were used for GIRAFFE, and the \feh{} from the {\it Gaia}-ESO Nice Node iDR5 analysis \citep[see ][for description]{Smiljanic2014, Worley2016} were used for UVES. 

\subsubsection{Outliers}
The sample was then investigated for outliers and discrepancies within this range of parameters, which are discussed below. 

\subsubsection*{Signal-to-noise}
\begin{figure}
\centering
\includegraphics[width=8.5cm]{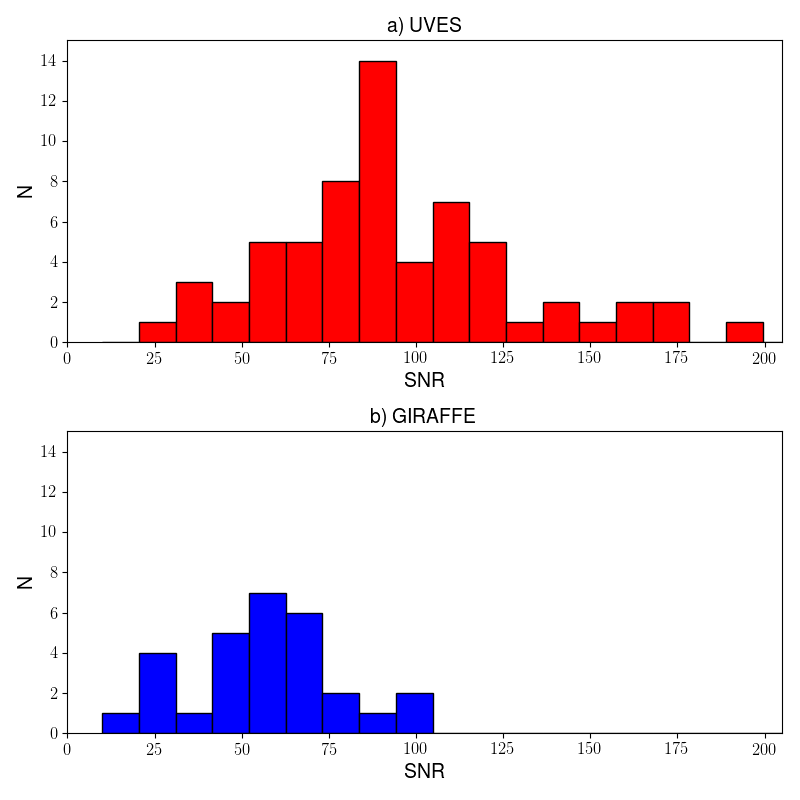}
\caption{S/N distribution for the K2 stars observed with a) UVES and b) GIRAFFE.}
\label{fig:snrhist}
\end{figure}

The signal-to-noise ratio (S/N) is a good indicator of the quality of the observed spectra. The distribution of the S/N for the UVES and GIRAFFE samples for the 90 K2 stars are shown in Fig.~\ref{fig:snrhist}. The majority of the spectra have S/N above 50. Those spectra with the lowest S/N may potentially suffer from insufficient signal causing deviations in the derived stellar parameters. This is considered in Sect.~\ref{sec:comp_nodes} in light of the stellar parameters determined by the two analysis teams.

\subsubsection*{IRFM not applicable}
There were three cases for which the magnitudes of the stars did not lie within the range of acceptable values to which the IRFM calibration relations can be applied. They are CNAMEs: 22072768-1440392, 22092416-0610474, and 22105015-1119135. The {\it preliminary} parameters were, however, complete for each and were thus used without an assessment against an IRFM \teff{}. 
Otherwise the median difference between the IRFM and {\it preliminary} \teff{} is $20\pm75$~K, showing good agreement in general. 

\subsubsection*{No iDR5 [Fe/H]}\label{sec:nofeh}
For the target with CNAME 22034179-0815421 and EPIC ID K2\_206298620, no recommended [Fe/H] was reported for iDR5, although \teff{} and \logg{} were provided. The radial velocity determination provided an associated \feh{} of $-2.55$, however the \teff{} associated with the radial velocity determination was greater than the iDR5 \teff{} by $\sim$300~K and the \logg{} was lower by $\sim$0.3. The IRFM \teff{} was in better agreement with the iDR5 \teff{}. An inspection of the spectrum using iSpec \citep{iSpec2014} was carried out, based on a comparison with synthetic spectra generated at both the iDR5 parameters and at the radial velocity determination parameters. In both cases, the \feh{} was re-derived, obtaining $-2.08$ and $-1.80,$ respectively. Based on this and the agreement with the IRFM \teff{}, the {\it preliminary} parameter estimate for this star was supplemented by \feh{}=$-2.08$ derived from the iDR5 parameters using iSpec. 

Based on the {\it preliminary} set being complete in $T_{\mathrm{eff}}$ and in reasonable agreement with the available photometric \teff{}, the {\it preliminary} \teff{} from spectroscopy were used to derive the {\it preliminary} \logg{} from seismology using the seismic analysis described in Paper~I carried out by the Birmingham team (hereafter referred to as BHAM). 

\subsection{Initial spectroscopic parameters}
The set of {\it preliminary} \teff{} and \feh{} compiled in Table~\ref{tab:ges_k2_prelim} are comprehensive but were unavoidably inconsistent in their source because not all of them were processed previously by the Gaia-ESO Survey. Thus, for the 90 stars found to have asteroseismic detections an initial analysis using just iSpec was carried out solely on these stars. For this analysis, the surface gravity was fixed to the {\it preliminary} seismic $\log g$ that was based on the {\it preliminary} \teff{}. This now homogeneous set from iSpec was used to explore any further inconsistencies between the iSpec spectroscopic and {\it preliminary} seismic results and fill in missing values as in Sect.~\ref{sec:nofeh}. The iSpec spectroscopic, \teff{} , was used to determine an associated seismic \logg{}. These comprise the {\it initial} set of parameters used as the starting point for the following iterative process between the two Gaia-ESO spectroscopic analyses and the seismic analysis. These {\it initial} parameters are listed in Table~\ref{tab:ges_k2_prelim}.



\section{Iterative determination of parameters}
\label{sect:parameters}
The goal of this process was to iterate between the spectroscopic effective temperature (\teff{}$_{\mathrm{,Spec}}$) and the seismic surface gravity ($\log g_{\mathrm{Seis}}$) to converge on a final set of independently-confirmed stellar parameters.
The seismic \logg{} was determined considering the parameters determined from seismology, namely the frequency of maximum power ($\nu_{\mathrm{max}}$) from the p-mode pulsation analysis and the spectroscopic parameter, \teff{}. We follow the scaling relation of:
\begin{equation}
    \log g = \log g_\odot + \log(\nu_{\mathrm{max}}/\nu_{\mathrm{max},\odot}) + \frac{1}{2}\log(T_{\mathrm{eff}}/T_{\mathrm{eff},\odot}),
\end{equation}
as in \cite{Morel2012}. Details of how this analysis works can be further found in \cite{morel14} for a sample of CoRoT targets and in \cite{apokasc2} for stars observed with {\it Kepler} and APOGEE. As discussed there, the seismic analysis of the p-modes is model-independent and so, the main source of uncertainty stems from the input temperature. Therefore, as the \teff{}$_{\mathrm{,Spec}}$ determination improves, the \logg{}$_{\mathrm{Seis}}$ determination also improves. 

We point to the discussion of \cite{morel14} that a change of 100~K in \teff{} only affects \logg{} by about 0.005. Therefore, significant improvement in \logg{} by, say, a change of the order of 0.1, requires a change in \teff{} that is much larger than typical uncertainties of \teff{}. Nonetheless, by fixing \logg{} to a value that is primarily only affected by $\nu_{\mathrm{max}}$ allows us to set a spectroscopic \teff{} scale that is consistent with the seismic log(g) and leads to improvement in other stellar quantities such as chemical composition, masses and ultimately ages (See also discussion in Paper I). 
 
\begin{figure}
\centering
\includegraphics[width=8cm]{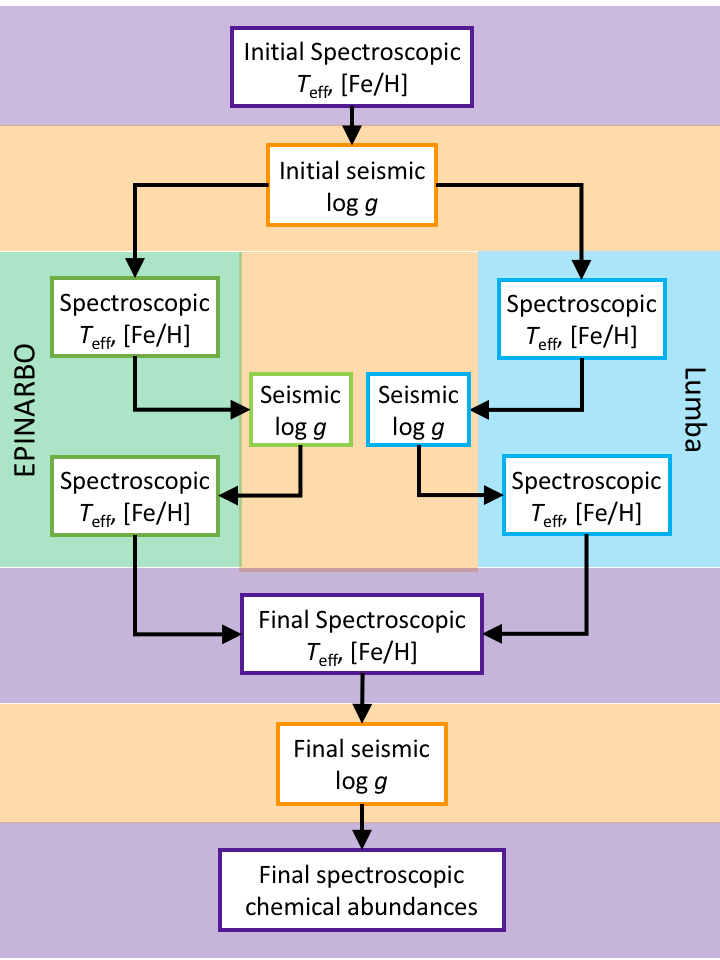}
\caption{Flow diagram of iterations between spectroscopic and seismic parameter determination.}
\label{fig:iterationflow}
\end{figure}

Two {\it Gaia}-ESO analysis nodes, EPINARBO and Lumba \citep[see][for further details of these and other nodes]{Smiljanic2014}, carried out the spectroscopic analysis of the GIRAFFE and UVES spectra of the K2 stars in the iterative procedure. The analysis methods are based on equivalent widths for EPINARBO and spectrum synthesis for Lumba (see Sects.~\ref{sec:epinarbo} and \ref{sec:lumba}). Following the requirements of the Gaia-ESO Survey, both methods use MARCS stellar atmosphere models \citep{Gustafsson2008} and the Gaia-ESO linelist \citep{heiterLL}. These two nodes were selected because they represent two of the most widely used methods for parameter determination in stellar spectroscopy \citep[equivalent widths and syntheses, see][]{Jofre18}. Figure~\ref{fig:iterationflow} illustrates the iterative process between spectroscopic and seismic parameter determinations that was followed in this analysis.

In summary, the initial spectroscopic parameters ($T_{\mathrm{eff,iniSpec}}$, [Fe/H]$_{\mathrm{iniSpec}}$) were used to determine the initial seismic $\log g_{\mathrm{iniSeis}}$. This set of $T_{\mathrm{eff,iniSpec}}$, [Fe/H]$_{\mathrm{iniSpec}}$ and $\log g_{\mathrm{iniSeis}}$ were given to both spectroscopic nodes. EPINARBO and Lumba were asked to fix the $\log g$ of their analysis to $\log g_{\mathrm{iniSeis}}$ and could otherwise use $T_{\mathrm{eff,iniSpec}}$ and [Fe/H]$_{\mathrm{iniSpec}}$ as priors in the re-determination of those values if needed. 

EPINARBO and Lumba each then returned a new set of parameters, $T_{\mathrm{eff,Spec1}}$ and [Fe/H]$_{\mathrm{Spec1}}$. Based on the $T_{\mathrm{eff,Spec1}}$ values, BHAM then calculated new $\log g_{\mathrm{Seis1}}$ values for each node. With their respective sets of $\log g_{\mathrm{Seis1}}$ values EPINARBO and Lumba once again determined the parameters based on the fixed $\log g_{\mathrm{Seis1}}$ to each provide $T_{\mathrm{eff,Spec2}}$ and [Fe/H]$_{\mathrm{Spec2}}$.

The two sets of $T_{\mathrm{eff,Spec2}}$ and [Fe/H]$_{\mathrm{Spec2}}$ were then combined to define the final set of spectroscopic parameters $T_{\mathrm{eff,finSpec}}$ and [Fe/H]$_{\mathrm{finSpec}}$. Based on $T_{\mathrm{eff,finSpec}}$ BHAM calculated the final seismic $\log g_{\mathrm{finSeis}}$.

These three values: $T_{\mathrm{eff,finSpec}}$, [Fe/H]$_{\mathrm{finSpec}}$ , and $\log g_{\mathrm{finSeis}}$ comprise the final stellar parameters of the K2@{\it Gaia}-ESO sample. In the final phase EPINARBO and Lumba were then asked to derive chemical abundances for each star based on these stellar parameters.

This was the defined procedure and goal of the K2@Gaia-ESO special project. However the homogenisation and combination of the results into a single final set of parameters per star required a detailed investigation of individual results. This was to ensure that each result was well-understood in an informed manner, which allows for reproducibility. We note that a careful homogenisation of the different node results has been a crucial focus of the {\it Gaia}-ESO Survey \citep{hourihanewg15,worleywg10}.

\subsection{Homogenisation strategy}
There were six sets of \teff{}, six sets of \feh{} and four sets of \logg{} produced in the iterative process. Within these sets are the high-resolution (UVES: 62 targets) and medium-resolution (GIRAFFE: 28 targets) subsamples. The high- and medium-resolution results were homogenised separately, as the lower resolution and smaller wavelength range of the GIRAFFE observations required more detailed quality assessment. The individual node results and the analysis undertaken to homogenise them are explained in the subsequent sections.

\subsection{EPINARBO analysis}\label{sec:epinarbo}
The EPINARBO analysis measures the equivalent widths (EW) with the DOOp code \citep{Doop2014}, which automatically measures equivalent widths with DAOSPEC \citep{daospec}. It then derives the stellar parameters and abundances with FAMA \citep{Magrini2014}, which calls spectrum synthesis code MOOG \citep{moog}.  The initial microturbulence parameter (\micro) was computed with the Gaia-ESO relation for stars with different \teff{} and \logg{} \citep{Smiljanic2014}.

For the K2@{\it Gaia}-ESO analysis, the surface gravity was fixed to the provided seismic value and EPINARBO iterated to converge on the equilibrium \teff{}, \feh{} and \micro. The stars for which the analysis
found a lack of sufficient Fe\,\textsc{i} and Fe\,\textsc{ii} lines were flagged by EPINARBO and, in particular, it was noted that the blended lines in the medium-resolution GIRAFFE spectra were not ideal for EW methods.

\begin{figure*}
\centering
\begin{minipage}[c]{\textwidth}
\centering
\centering
\includegraphics[width=18.0cm]{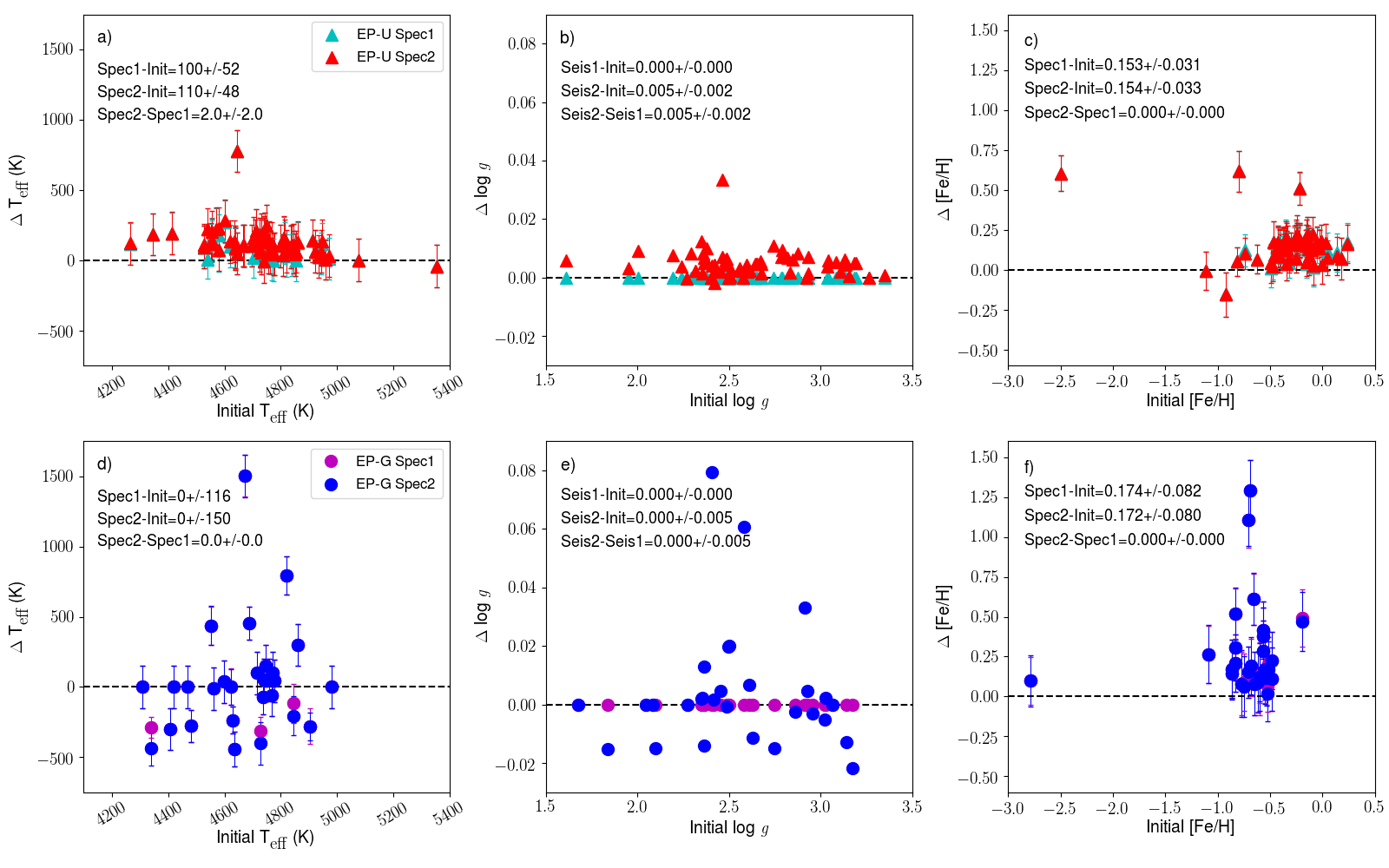}
\caption{Stellar parameter iterations for EPINARBO a-c) UVES sample; d-f) GIRAFFE sample. For \teff{}, \logg{} and \feh{}, comparison of initial parameters against Spec1 and Spec2. The median of differences and MAD values are specified.}
\label{fig:ep_giruves_iter}

\includegraphics[width=18.0cm]{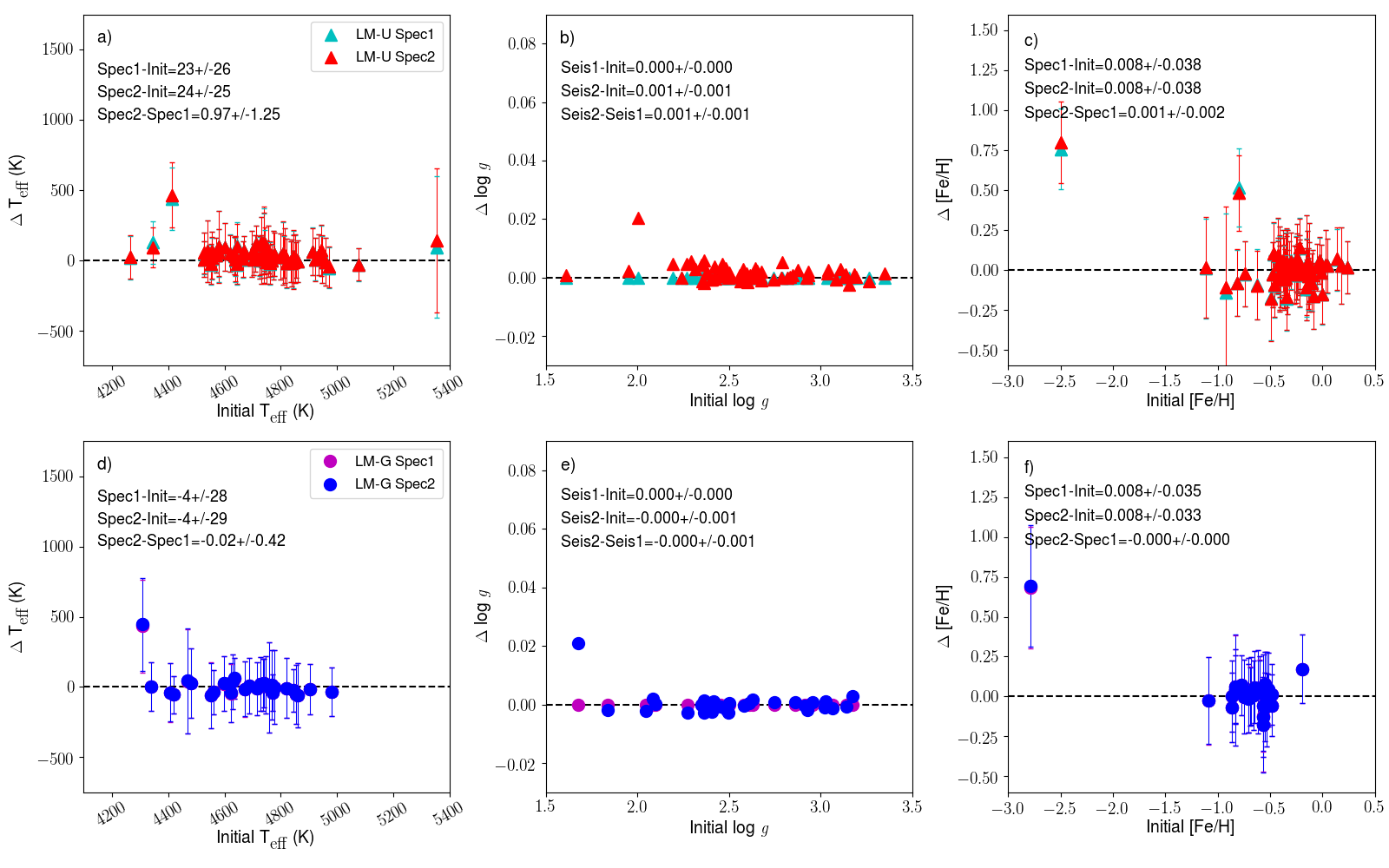}
\caption{Stellar parameter iterations for Lumba a-c) UVES sample; d-f) GIRAFFE sample. For \teff{}, \logg{} and \feh{}, comparison of initial parameters against Spec1 and Spec2. The median of differences and MAD values are specified.}\label{fig:lm_giruves_iter}
\end{minipage}
\end{figure*}

Figure~\ref{fig:ep_giruves_iter} shows the progression of the EPINARBO results iteration. Figure~\ref{fig:ep_giruves_iter}a-c compares the initial parameters ($T_{\mathrm{eff,iniSpec}}, \log g_{\mathrm{iniSeis}}, $\feh{}$_{\mathrm{iniSpec}}$) to each iterated set of parameters (red: $T_{\mathrm{eff,Spec1}}, \log g_{\mathrm{Seis1}}, $\feh{}$_{\mathrm{Spec1}}$; blue: $T_{\mathrm{eff,Spec2}}, \log g_{\mathrm{Seis2}}, $\feh{}$_{\mathrm{Spec2}}$) for the UVES analysis. Figure~\ref{fig:ep_giruves_iter}d-f are the same but for GIRAFFE.

The median and median absolute difference (MAD) of the difference between iteration sets for each parameter and each iteration are also shown. There is little variation from Spec1 to Spec2 for both UVES and GIRAFFE. GIRAFFE shows more scatter in the results while the UVES results seem more stable. The plots of seismic \logg{} are included for completeness showing that any large variation in \teff{} between iterations does not result in much variation in the \logg{}. This agrees with the findings of \cite{morel14}.  See Sect.~\ref{sec:comp_nodes} for more discussion on this.

\begin{figure*}[!ht]
\centering
\includegraphics[width=18cm]{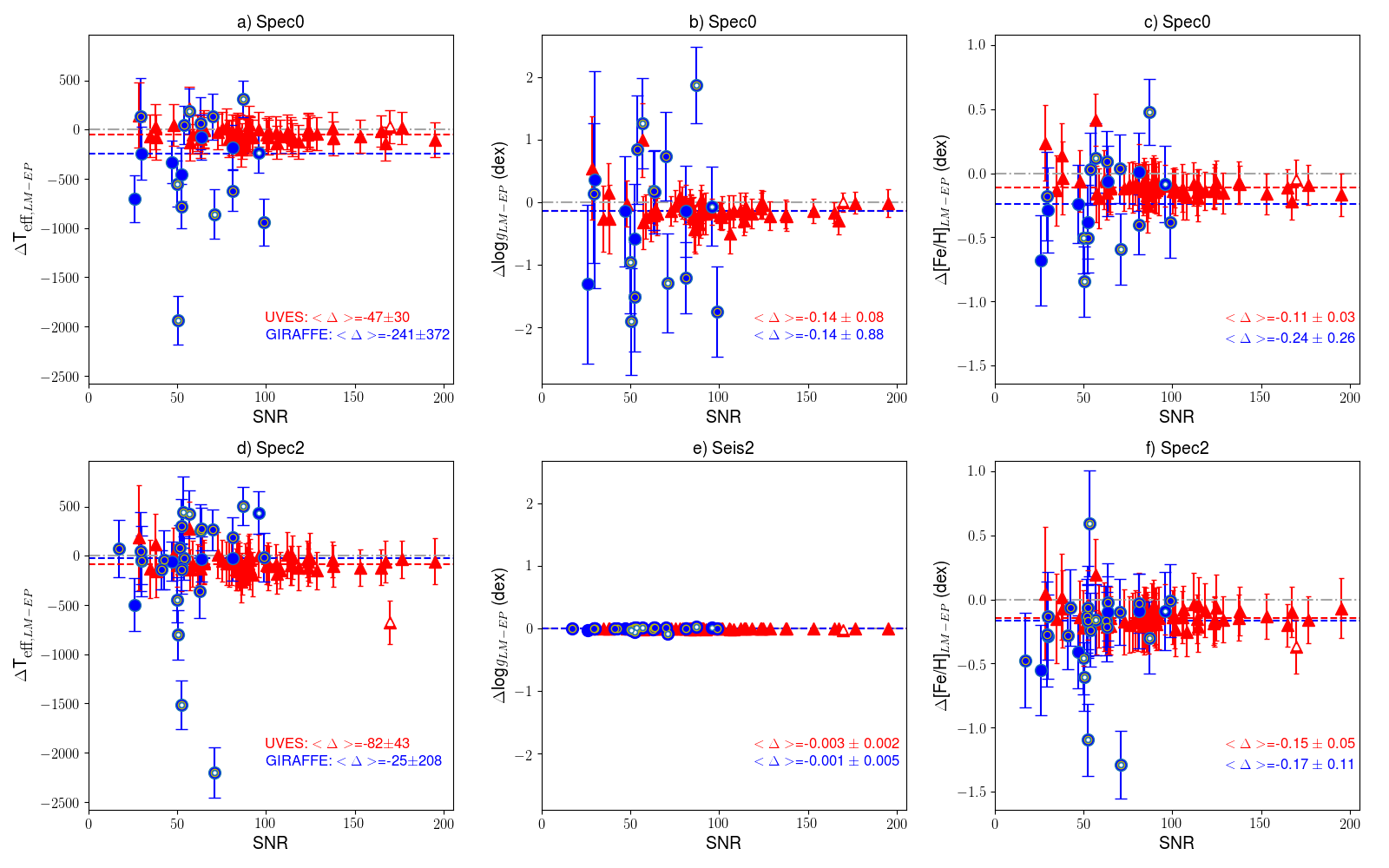}
\caption{Comparison between the EPINARBO and Lumba parameters against signal-to-noise (S/N) determined for Spec0 in the top row and Spec2 in the bottom row. Left to right, the panels compare spectroscopic \teff{}, spectrocopic/seismic \logg,{} and spectroscopic \feh,{} in turn, for UVES (red) and GIRAFFE (blue) spectra. Yellow circles are stars rejected by EPINARBO for inconsistent Fe\,\textsc{i} and Fe\,\textsc{ii} abundances in Spec0. Central white circles (GIRAFFE) or white triangle (UVES) are stars rejected by EPINARBO for inconsistent Fe\,\textsc{i} and Fe\,\textsc{ii} abundances in Spec2. The median and MAD of the difference is given for each.}
\label{fig:spec0_spec2}
\end{figure*}%



\subsection{Lumba analysis}\label{sec:lumba}
The Lumba analysis \citep{gavellumba} performs spectrum synthesis using Spectroscopy Made Easy \citep[SME:][]{Valenti1996,Piskunov2017}. For the K2@{\it Gaia}-ESO analysis, the surface gravity was fixed to the provided seismic value and Lumba iterated to converge on the equilibrium \teff{}, \feh,{} and \micro{} \citep{Smiljanic2014}.

Figure~\ref{fig:lm_giruves_iter} shows the iterative process for the Lumba UVES and GIRAFFE analyses as for Fig.~\ref{fig:ep_giruves_iter}. There are some distinct outliers for each instrument set, however, the Lumba results are generally very stable between iterations as expected \citep{morel14}.

\subsection{Comparison of node parameters}\label{sec:comp_nodes}
Within the {\it Gaia}-ESO Survey, all of the nodes perform a fully spectroscopic analysis of the {\it Gaia}-ESO spectra. While the determination of unconstrained spectroscopic \teff{}, \logg,{} and \feh{} was not the goal of this study, it is interesting to compare the unconstrained spectroscopic parameters to those determined by iteration between spectroscopy and astereoseismology. For the purposes of this paper, the unconstrained spectroscopic \teff{}, \logg,{} and \feh{} determined by each node are referred to as Spec0.

Figure~\ref{fig:spec0_spec2} directly compares the EPINARBO and Lumba parameters derived for Spec0 (top row) and for the final iteration, Spec2 (bottom row), for GIRAFFE (blue), and UVES (red), respectively. Based on Fig.~\ref{fig:lm_giruves_iter} and Fig.~\ref{fig:ep_giruves_iter}, there was little movement between iterations for each node. Therefore inspecting the final parameters (Spec2) from each node was deemed sufficient. The median and MAD for each parameter between the two nodes are also given. The greater spread of the difference in the GIRAFFE parameters compared to the UVES parameters is clearly seen. The UVES results are in good agreement between the nodes for both Spec0 and Spec2.

As stated in Sect.~\ref{sec:epinarbo}, EPINARBO reported that the EW method found inconsistencies in Fe\,\textsc{i} and Fe\,\textsc{ii} abundances for the GIRAFFE spectra due to the blending of spectral features at that resolution and, indeed, a total of nine stars are not included at all for GIRAFFE Spec0 as the EW method did not converge on a result. The particular stars with converged results but that were flagged by EPINARBO are highlighted in all panels in Fig.~\ref{fig:spec0_spec2} as data points with a yellow circle as those rejected by EPINARBO in Spec0, and data points with central white dots (GIRAFFE), or white triangles (UVES) as those rejected by EPINARBO in Spec2. The use of the EPINARBO flagged stars for assessing the quality of the results is explained in detail in Sect.~\ref{sec:node_qual}.

For Spec0 even those that are not rejected have a wide spread, while for Spec2 the stars that are not rejected are tightly distributed and in good agreement with the Lumba results. Certainly this comparison shows that for the GIRAFFE spectra (lower resolution and smaller wavelength range that UVES) fixing the \logg{} using astereoseismology has allowed the spectroscopic \teff{} to be better constrained and, thus, there is better agreement between the methods and more of the sample is available for abundance analysis. The high-resolution and greater wavelength range of the UVES sample produces good agreement between the nodes results for both Spec0 and Spec2.

The differences in node parameters in Fig.~\ref{fig:spec0_spec2} are shown against S/N. There is no obvious indication that as the S/N decreases, the differences between the node parameters increase. Indeed, for the UVES sample, the spread in the differences is fairly consistent and minimal across the S/N range for both Spec0 and Spec2. For the GIRAFFE sample, there is a large scatter generally for this smaller sample of 28 stars. As there was no obvious trend with S/N, the differences between the node parameters were used directly in the assessment of the quality of the results.

Considering the Spec2 results in particular, which are the set from which the final stellar parameters will be determined, there is, overall, an offset in \feh{} between the nodes for both the UVES and GIRAFFE analyses ($\Delta$ \feh{} $\simeq -0.15$). There is an offset in \teff{} ($\Delta$ \teff{}$ \simeq -80$~K). This is discussed further in Sect.~\ref{sec:finuvesparam}.

\begin{figure}
\centering
\includegraphics[width=8.5cm]{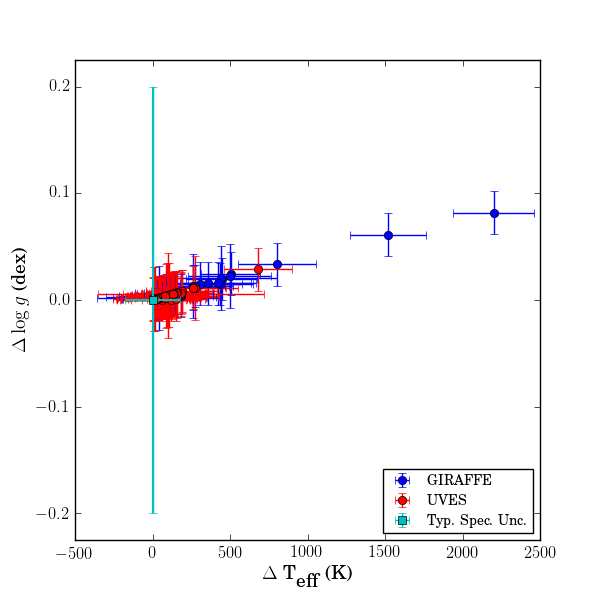}
\caption{Node difference in final seismic \logg{} against the difference in final \teff{} for the high (red) and medium (blue) resolution data. Typical spectroscopic uncertainties are shown in cyan.}
\label{fig:dteff_dlogg}
\end{figure}

The \logg{} in all cases are those from the seismic analysis, based on the respective spectroscopic \teff{}. There are some high discrepancies found between the nodes ($\Delta$ \teff{}$ > 500$~K, $\Delta$ \feh{} $> 0.5$) for certain stars, most particularly in the GIRAFFE analysis reflected in the high value of the MAD (MAD$_{T_{\mathrm{eff}}} > 308$~K, MAD$_{[Fe/H]} > 0.15$) compared to the tighter agreement for the UVES analyses (MAD$_{T_{\mathrm{eff}}} > 60$~K, MAD$_{[Fe/H]} > 0.07$). The largest disagreement in \teff{} equates to a very small shift in seismic \logg{} ($\Delta$ \teff{}$ \approx -2200$~K corresponds to~$\Delta$ \logg{} $\approx -0.09$). 

Figure~\ref{fig:dteff_dlogg} shows the difference between the two sets of node results for the final seismic \logg{} against the difference in final spectroscopic \teff{} upon which the final seismic \logg{} values are based. In cyan we show the typical uncertainties for \teff{} and \logg{} when determined spectroscopically.  Differences in spectroscopic \teff{} greater than 500~K equate to less than a 0.1 difference in seismic \logg,{} which agrees with the discussion in Sect.\ref{sect:parameters}. While asteroseismology pinpoints the \logg{}, it does so from a large potential range in \teff{}. Therefore, complementary methods are needed, such as spectroscopy, to accurately converge on all stellar parameters.

The goal at this point was to combine the EPINARBO and Lumba Spec2 results to produce a final spectroscopic \teff{} (and \feh{}) from which a final seismic \logg{} could be calculated. However, prior to this, it was important to understand the differences between the node analyses, particularly with regard to stars for which there was large disagreement, as we did not want to blindly assume that a mean of the parameters from the two nodes was sufficient as a best final value.
To assess the goodness of the results, ancillary information was compared to the differences in the parameters to ensure that the results that were selected were of the best quality possible.

Information that was considered was: 1) Node uncertainties on parameters; 2) Quality assessment reported by Nodes; 3) Microturbulence (\micro{}); 4) Normalised $\chi^2$ between observed and synthetic spectra; and 5) Comparison to photometric \teff{}.

As stated above, there was no obvious trend with S/N so it was not considered in the following quality assessment.
The IRFM and Node final parameters are given in Table~\ref{tab:uves-gir_outliers} for the UVES and GIRAFFE samples.

\subsubsection{Uncertainties on parameters}
\begin{figure*}
\begin{minipage}[b]{18cm}
\includegraphics[width=\textwidth]{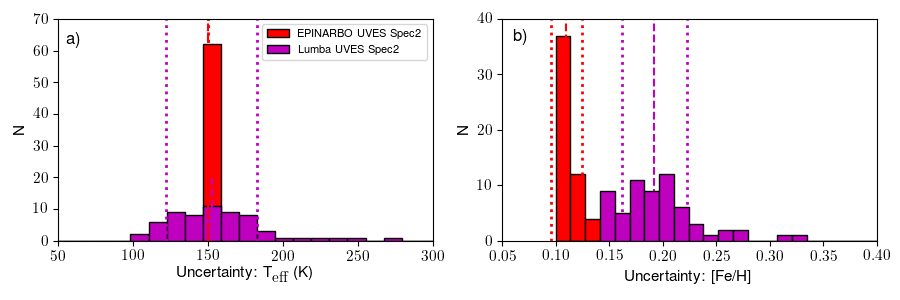}
\end{minipage}
\begin{minipage}[b]{18cm}
\includegraphics[width=\textwidth]{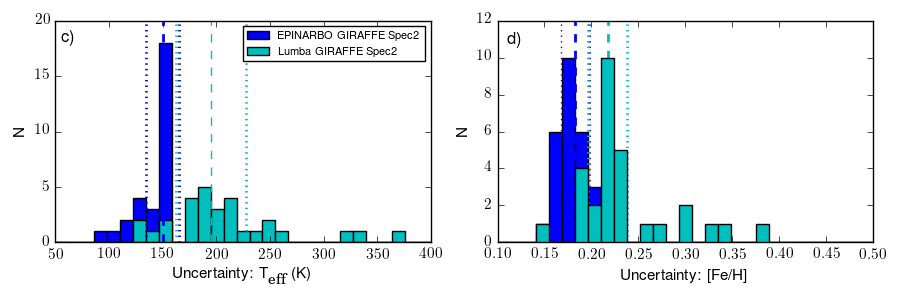}
\end{minipage}
\caption{Uncertainty distribution for \teff{} and \feh{} for EPINARBO and Lumba for UVES (a \& b), and GIRAFFE (c \& d). The median for each distribution is shown as a dashed line with the respective colour, $\pm\sigma$ is shown as dotted lines.}
\label{fig:eplm_giruves_errors}
\end{figure*}

Figure~\ref{fig:eplm_giruves_errors} shows the uncertainty distributions for the spectroscopic parameters \teff{} and \feh{} for both EPINARBO and Lumba for the GIRAFFE and UVES analyses. There is a systematic offset between the node distributions and EPINARBO reports a tighter distribution (with strict upper limit for \teff{}) than Lumba for both the UVES and GIRAFFE samples.

Each node has reported the uncertainties as best suits their analysis pipelines and, thus, they are internally consistent. However, the uncertainties are not calculated in the same way between the nodes. Due to this difference, they cannot be used to compare the node results (i.e. as a weight) to determine the best value for individual stars.

The difficulty of making comparisons between node results based on the respective uncertainty distributions, when the uncertainties are not defined in a standard way, has been a challenge throughout the lifetime of the {\it Gaia}-ESO Survey and for studies in Galactic Archeaology in general. Further discussions on uncertainty analyses can be found in \cite{Jofre18}.

\subsubsection{Quality assessment}\label{sec:node_qual}
EPINARBO provided a list of stars for which the sample of Fe\,\textsc{i} and Fe\,\textsc{ii} lines was inadequate for a robust analysis. These are indicated with a '*' in the {\it CNAME} column of Table~\ref{tab:uves-gir_outliers}. There are eight of these stars within the GIRAFFE analysis and 1 star within the UVES analysis.

The uncertainty distribution of the Lumba results shows high tails inferring less confidence in the parameter determination for those stars. Objects with \teff{}~uncertainty~$> 294$~K (244~K; median \teff{} uncertainty + 3$\sigma$) and \feh{} uncertainty~$> 0.28$ (0.28; the median [Fe/H] uncertainty + 3$\sigma$) were assumed to indicate lower confidence by Lumba in the parameters for the GIRAFFE (UVES) samples.

\subsubsection{Microturbulence (\micro{})}
The \micro{} was provided by the nodes for both the GIRAFFE and UVES analyses. The \micro{} relation derived based on iDR1 was used often as a starting point for iterations or as a derived value depending on the respective node procedures. The difference between node \micro{} values were in some cases very high, particularly for the GIRAFFE analysis.

\begin{figure}
\centering
\includegraphics[width=9cm]{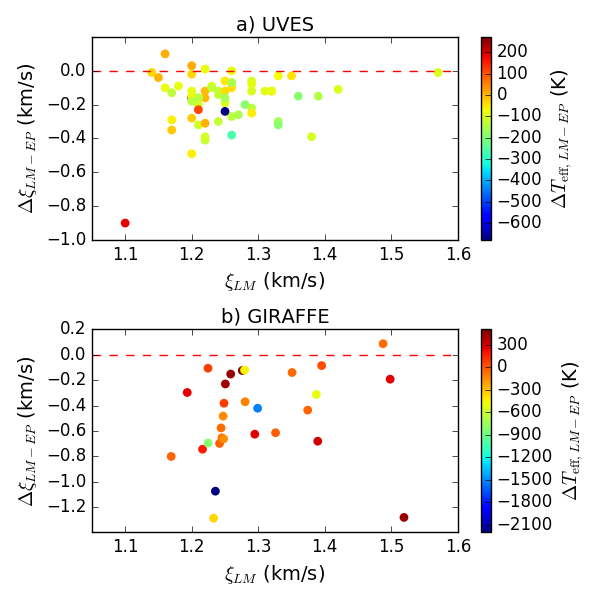}
\caption{$\Delta$\micro{}$_{LM-EP}$ against \micro{}$_{LM}$ with a colourmap of $\Delta T_{\mathrm{eff},LM-EP}$ for a) UVES; b) GIRAFFE.}
\label{fig:xi_comp}
\end{figure}

Figure~\ref{fig:xi_comp} compares the node \micro{} values for both the a) UVES and b) GIRAFFE analyses directly with a colourmap of $\Delta T_{\mathrm{eff},LM-EP}$. For UVES, there was one star with a large difference in \micro{}, but otherwise there was no strong evidence of $\Delta$\micro{} correlating with $\Delta T_{\mathrm{eff}}$.
For GIRAFFE the extreme differences in \micro{} typically were accompanied by other extremes in parameters which are described below. A difference of \micro{}$ > 1$~km/s was used as a threshold when considered alongside the other indicators of goodness of fit.

This comparison of the \micro{} values between the two nodes and the two resolutions for the K2@{\it Gaia}-ESO sample revealed an error in the microturbulence relation made available for iDR1. Thanks to this work, the relation was re-derived based on iDR5 values for the iDR6 analysis \citep[see][]{worleywg10}.

\subsubsection{Normalised $\chi^2$}
It was necessary to independently assess the goodness of fit of the node solutions for the GIRAFFE analyses due to large differences between the reported node parameters. To this end, synthetic spectra for both HR10 and HR21 wavelength ranges were generated for each star using Turbospectrum \citep{plez2012} and the MARCS stellar atmosphere models \citep{Gustafsson2008}. Interpolating between the models \citep{MasseronPhD}, the spectra were generated at the parameters derived by each node and then a normalised $\chi^2$ was calculated between the synthetic spectrum and the observed spectrum. The observed spectrum was normalised to the synthetic spectrum in each case. In this way, the fit was optimised to the solution of each node. The goal was to look for obvious discrepancies in $\chi^2$ to discard extreme outliers.

\begin{figure}
\centering
\includegraphics[width=9cm]{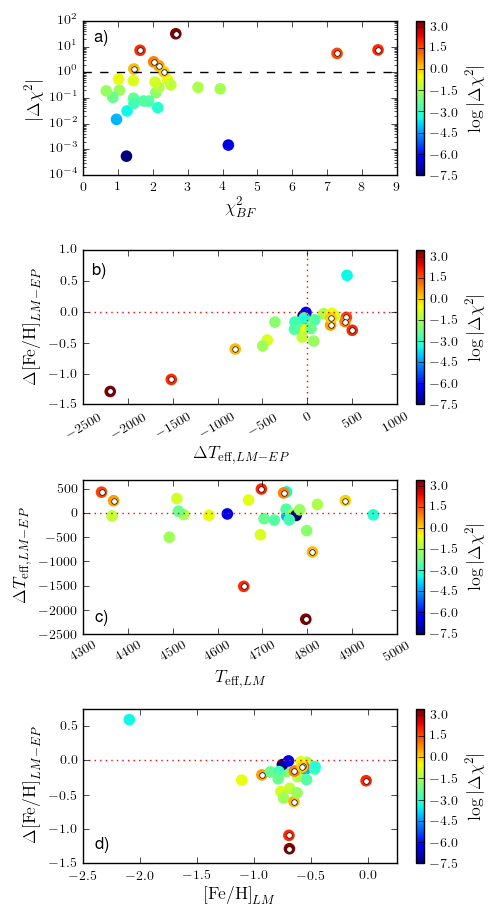}
\caption{Assessment of rejection criteria based on absolute difference of $\chi^2$ between EPINARBO and Lumba. In each case the colourmap is $\log|\Delta\chi^2|$: a) $|\Delta\chi^2|$ in a log scale against the best fit (BF) $\chi^2$. Rejection limit at 1.0 is shown as a dashed line; b) $\Delta$\feh{}$_{LM-EP}$ against $\Delta$\teff {}$_{LM-EP}$; c) $\Delta$\teff{}$_{,LM-EP}$ against \teff{}$_{,LM}$; d) $\Delta$\feh{}$_{LM-EP}$ against \feh{}$_{LM}$. The data points with a white centre are those rejected when $|\Delta\chi^2| > 1.0$ }
\label{fig:gir_chi2}
\end{figure}

Two measures were used for assessing the goodness indication of the $\chi^2$: a) absolute difference between the EPINARBO $\chi^2$ and the Lumba $\chi^2$ ($|\Delta\chi^2|$); b) absolute values of each node $\chi^2$. 

The set of parameters that represent the best-fit by $\chi^2_{BF}$ between EPINARBO and Lumba was used as the comparison set to avoid assuming that either EPINARBO or Lumba parameters were the best. Figure~\ref{fig:gir_chi2}a compares $|\Delta\chi^2|$ with $\chi^2_{BF}$ also showing $\log|\Delta\chi^2|$ as a colourmap. This $\log|\Delta\chi^2|$ colourmap is used in Fig.~\ref{fig:gir_chi2}b-d to explore the trend with other goodness of fit indicators. When considered in combination with other indicators of goodness of fit, typically all $|\Delta\chi^2| > 1.0$ indicated poor agreement between the nodes and so, in these cases, the node parameter set with the highest $\chi^2$ value was rejected. This threshold is shown as a dashed line in Fig.~\ref{fig:gir_chi2}a and the rejected points have a white dot at the centre.

Figure~\ref{fig:gir_chi2}b compares the difference in \feh{} between the nodes with the difference in \teff{} between the nodes. Typically, when one is large, the other is large, and the $\log|\Delta\chi^2|$ is also large, as   expected. Figures~\ref{fig:gir_chi2}c and d explore whether there is any trend of $\log|\Delta\chi^2|$ and $\Delta$\teff{}$_{,LM-EP}$ with \teff{}$_{,LM}$, and any trend of $\log|\Delta\chi^2|$ and $\Delta$\feh{}$_{LM-EP}$ with \feh{}$_{LM}$, in case a particular part of the parameter space was particularly susceptible. The Lumba \teff{} and \feh{} were used as reference. There were no obvious trends in either case.

\subsubsection{Photometric \teff{}}
The GIRAFFE sample  showed, in particular, large disagreements between the two sets of node results. A comparison to the photometric \teff{} provided another useful indicator as the photometric and spectroscopic \teff{} should be in relatively good agreement. Interstellar extinction is not expected to play a significant role in the photometric colours as these stars are towards the Galactic pole. When considered in combination with the other indicators, a difference of \teff{}$ > 250$~K between the photometric and spectroscopic \teff{} was empirically defined as being too great.

\subsubsection{Final GIRAFFE spectroscopic parameters}
The final spectroscopic stellar parameters (\teff{}, \feh{}, \micro{}) of the 28 stars in the GIRAFFE sample were defined by first inspecting the node results using the criteria as described above to reject any poor parameter determinations (see Table~\ref{tab:uves-gir_outliers} for the photometric and node parameters, and the outlier assessment, as well as Table~\ref{tab:fin_stellar_params} for the final spectroscopic parameters). In these cases, the remaining node parameter set was used as indicated in column 4 of Table~\ref{tab:fin_stellar_params}. There were 12 cases where only one of the node parameter sets was used. For eight of these cases, EPINARBO had reported large discrepancies between the Fe\,\textsc{i} and Fe\,\textsc{ii} abundances, or a lack of suitable Fe\,\textsc{ii} lines, for these stars. The $\chi^2$ test confirms these as poor fits to the data, with the Lumba node providing significantly better fitting results. 

The remaining four cases were considered as follows:
\begin{enumerate}
\item CNAME 22082566-1532383: The large discrepancies between the EPINARBO \teff{} and the IRFM \teff{} (--288~K), and also the Lumba \teff{} (--360~K), considered with the large discrepancy between the EPINARBO \micro{} and the Lumba \micro{}  (--1.29~km/s) indicated that there were issues with the EPINARBO result and so the Lumba result was taken as the final parameter set.
\item CNAME 22114679-1126477: Similarly, the large discrepancies between the EPINARBO \teff{} and the IRFM \teff{} (--455~K), and also the Lumba \teff{} (--497~K), considered with the large discrepancy between the EPINARBO \feh{} and the Lumba \feh{} (--0.55) indicated that there were issues with the EPINARBO result and so, the Lumba result was taken as the final parameter set.
\item CNAME 22032304-0754111: The large discrepancy between the EPINARBO $\chi^2$ and the Lumba $\chi^2$ (-1.81) indicated an issue with the fit to the data; in this case, EPINARBO had the worst fit. This was borne out by inspecting the individual HR10 and HR21 $\chi^2$ values, for which the HR21 $\chi^2$ values were comparable between the nodes, but the EPINARBO $\chi^2$ for HR10 was significantly worse. The difference in \teff{} reflected this. The Lumba result was taken as the final parameter set.
\item CNAME 22154067-0627110: The discrepancy above the threshold of 1.0 between the EPINARBO $\chi^2$ and the Lumba $\chi^2$ (-1.06) indicated an issue with the fit to the data; also in this case EPINARBO had the worst fit. Inspection of the individual HR10 and HR21 $\chi^2$ values also showed that the HR21 $\chi^2$ values were comparable between the nodes, but the EPINARBO $\chi^2$ for HR10 was significantly worse. The difference in \teff{} reflected this. The Lumba result was therefore taken as the final parameter set.
\end{enumerate}

\subsubsection{Final UVES spectroscopic parameters}\label{sec:finuvesparam}
The final spectroscopic stellar parameters (\teff{}, \feh{}, \micro{}) of the 62 stars in the UVES sample were defined also by first inspecting the node results using the criteria as described above to reject any poor parameter determinations. Overall, there was much better agreement between Lumba and EPINARBO for the UVES sample than for the GIRAFFE sample and so, it was not necessary to derive the independent $\chi^2$ as an additional measure of goodness of fit. The difference in \teff{} and \feh{} between nodes is seen in Fig.~\ref{fig:spec0_spec2}. While there is a mean difference that is comparable to the scatter, we are not in a position to favour one node over the other since we do not have reference values for stars of this sample. For {\it Gaia}-ESO, both nodes have been independently calibrated with the common set of reference parameters of the {\it Gaia} FGK benchmark stars. Therefore, it is not a straightforward task to decide which of the nodes in this particular case, where the pipelines were adapted to iterate with seismology, might be less accurate. In this light, the average of the node results were taken as the final spectroscopic parameters.
See Table~\ref{tab:uves-gir_outliers} for the photometric and node parameters and the outlier assessment, as well as Table~\ref{tab:fin_stellar_params} for the final spectroscopic parameters.

There are two remaining cases where the differences are significantly larger and are considered as follows:
\begin{enumerate}
\item CNAME 22000793-1203412: The large discrepancies between the Lumba \teff{} and the IRFM \teff{} (--263~K) considered with the large discrepancy between the EPINARBO \micro{} and the Lumba $\xi$  (--0.90~km/s) and the very large uncertainty on the Lumba \teff{} (512~K) and \feh{} (0.50) indicated that there were issues with the Lumba result and so the EPINARBO result was taken as the final parameter set.
\item CNAME 22032202-0829154: EPINARBO flagged this star as the only UVES star that they found to have a large discrepancy between the abundances derived from the Fe\,\textsc{i} and Fe\,\textsc{ii} lines. The EPINARBO and Lumba \teff{} disagree by $\Delta T_{\mathrm{eff},LM-EP}=-679$~K. Both do not agree well with the IRFM \teff{} with differences of -296~K and 383~K respectively. The initial spectroscopic \teff{} derived from iSpec was 4643~K, which also does not agree with the IRFM \teff{} nor EPINARBO. There is, however, no substantial difference between the seismic \logg{} derived for each node (0.03). As EPINARBO flags an issue between the Fe\,\textsc{i} and Fe\,\textsc{ii,} the Lumba parameters were taken as the final set. In the subsequent chemical analysis, the final Fe\,\textsc{i} and Fe\,\textsc{ii} are in good agreement (Fe\,\textsc{i}-Fe\,\textsc{ii}=0.1), confirming this choice of parameters.
\end{enumerate}

\subsection{Final K2@{\it Gaia}-ESO stellar parameters}
The homogenisation process carried out above resulted in the final catalogue of stellar parameters for the 90 K2@{\it Gaia}ESO red giant stars. These are shown in Fig.~\ref{fig:initialHRD}, with the \feh{} as a colourmap. The final stellar parameters are list in Table~\ref{tab:fin_stellar_params} and are provided in the final K2@{\it Gaia}-ESO catalogue of parameters and abundances which is available online. The columns of the full catalogue are listed in Table~\ref{tab:onelinetab_cols}.

\section{K2@{\it Gaia}-ESO chemical abundances}
\label{sect:abundances}
The final phase of the analysis was the measurement of chemical abundances for those elements typically measured for {\it Gaia}-ESO by each node. Based on the final set of stellar parameters (\teff{}, \logg{}, \feh{}, and \micro{}), each node determined their typical chemical abundances for the GIRAFFE and UVES spectra. The nodes provided abundances per star and also abundances per spectral line. 

The homogenisation was undertaken using the line-by-line abundances in order for the results to be combined in a more informed way such that the effect of each spectral feature is considered separately. Even with the detailed work required to produce a refined, high-quality line list \citep{heiterLL}, some of the spectral features used here are less well-modelled than others, which can be due to less robust line information or a strong influence coming from surrounding features, or less robust normalisation of the continuum about that feature. The larger pool of values for each spectral feature across the two nodes compared to a single global value from each node allows for the outlying spectral features to be identified and discarded, if needed, and for features that are in good agreement between the nodes to be more highly weighted.  

This was carried out only for those elements for which at least one of the nodes measured more that 50\% of the sample, that is, more than 14 stars for GIRAFFE and more than 31 stars for UVES. The threshold of 50\% was chosen to meet the requirement for a reference set to be as complete as possible in the reported measurements across all the stars in the set. The node results for any element with less than these thresholds were discarded. The remaining elements and number of K2 stars that were measured by each node and the median number of spectral lines each node measured per element are listed in Table~\ref{tab:abund_sum}.  

\setcounter{table}{3}
\begin{table}[htbp]
  \centering
  \footnotesize
  \caption{List of elements measured for each element by EPINARBO and Lumba for the GIRAFFE and UVES K2 stars. The number of stars per element and the median number of clean spectral lines measured per element are given.}\label{tab:abund_sum}
    \begin{tabular}{c|cc|cc|cc|cc}
    \hline\hline
          & \multicolumn{4}{c|}{UVES}      & \multicolumn{4}{c}{GIRAFFE} \\
          & \multicolumn{2}{c}{EPINARBO} & \multicolumn{2}{c|}{Lumba} & \multicolumn{2}{c}{EPINARBO} & \multicolumn{2}{c}{Lumba} \\
          \hline
    Element & Stars & Lines & Stars & Lines & Stars & Lines & Stars & Lines \\
    \hline
    Na\,\textsc{i}   & 61    & 2     & 62    & 1     & -     & -     & -     & - \\
    Mg\,\textsc{i}   & 62    & 2     & 62    & 2     & 28    & 2     & 28    & 3 \\
    Al\,\textsc{i}   & 61    & 2     & 62    & 3     & 28    & 2     & 24    & 1 \\
    Si\,\textsc{i}   & 62    & 6     & 62    & 8     & 28    & 2     & 25    & 4 \\
    Si\,\textsc{ii}   & 44    & 1     & 62    & 2     & -     & -     & -     & - \\
    Ca\,\textsc{i}   & 62    & 6     & 62    & 19    & 21    & 1     & -     & - \\
    Ca\,\textsc{ii}   & -     & -     & 61    & 2     & -     & -     & -     & - \\
    Sc\,\textsc{i}   & 61    & 4     & 62    & 4     & -     & -     & -     & - \\
    Sc\,\textsc{ii}   & 62    & 4     & 62    & 4     & 24    & 1     & 28    & 1 \\
    Ti\,\textsc{i}   & 62    & 17    & 62    & 39    & 28    & 10    & -     & - \\
    Ti\,\textsc{ii}   & 62    & 7     & 62    & 7     & 28    & 2     & -     & - \\
    V\,\textsc{i}    & 62    & 14    & 62    & 24    & 20    & 2     & -     & - \\
    Cr\,\textsc{i}   & 62    & 12    & 62    & 20    & -     & -     & 25    & 1 \\
    Cr\,\textsc{ii}   & -     & -     & 45    & 1     & -     & -     & -     & - \\
    Mn\,\textsc{i}   & -     & -     & 62    & 10    & -     & -     & 27    & 3 \\
    Fe\,\textsc{i}   & 62    & 64    & 62    & 14    & 28    & 9     & 28    & 34 \\
    Fe\,\textsc{ii}   & 62    & 8     & 62    & 3     & 28    & 2     & -     & - \\
    Co\,\textsc{i}   & 62    & 13    & 62    & 18    & 28    & 3     & 27    & 3 \\
    Ni\,\textsc{i}   & 62    & 13    & 62    & 24    & 28    & 2     & 22    & 2 \\
    Cu\,\textsc{i}   & 62    & 1     & 62    & 2     & -     & -     & -     & - \\
    Zn\,\textsc{i}   & 61    & 2     & 62    & 1     & -     & -     & -     & - \\
    Y\,\textsc{ii}    & 62    & 6     & 62    & 5     & 27    & 1     & -     & - \\
    Zr\,\textsc{i}   & 59    & 5     & 62    & 5     & -     & -     & -     & - \\
    Zr\,\textsc{ii}   & 34    & 1     & -     & -     & -     & -     & -     & - \\
    Ba\,\textsc{ii}   & 53    & 1     & 62    & 3     & -     & -     & -     & - \\
    La\,\textsc{ii}   & 60    & 3     & -     & -     & -     & -     & -     & - \\
    Ce\,\textsc{ii}   & 60    & 2     & -     & -     & -    & -     & -     & - \\
    Nd\,\textsc{ii}   & -     & -     & 62    & 8     & -     & -     & -     & - \\
    Eu\,\textsc{ii}   & 57    & 1     & 62    & 1     & -     & -     & -     & - \\
    \hline
    \end{tabular}%
  \label{tab:listelements}%
\end{table}%

The compilation of spectral lines used for {\it Gaia}-ESO is described in \cite{heiterLL} and the individual sources for the spectral lines for each element, summarised in \cite{heiterLL}, are as follows: Na\,\textsc{i} - \citet{NIST10}; Mg\,\textsc{i} - \citet{NIST10}; Al\,\textsc{i} - \citet{WSM}; Si\,\textsc{i} - \citet{GARZ, K07}; Si\,\textsc{ii} - \citet{K12}; Ca\,\textsc{i} - \citet{DIKH, K07, Sm, SN, S0, SR}; Ca\,\textsc{ii} - \citet{TB}; Sc\,\textsc{i} - \citet{LD}; Sc\,\textsc{ii} - \citet{K09, LD}; Ti\,\textsc{i} - \citet{K10, NWL, LGWSC}; Ti\,\textsc{ii} - \citet{WLSC, K10, RHL}; V\,\textsc{i} - \citet{K09}; Cr\,\textsc{i} - \citet{WLHK, K10, SLS}; Cr\,\textsc{ii} - \citet{SLd, K10, PGBH}; Mn\,\textsc{i} - \citet{K07, DLSSC}; Fe\,\textsc{i} - \citet{K07, FMW, BKK, BWL, BK0}; Fe\,\textsc{ii} - \citet{BSScor, K13}; Co\,\textsc{i} - \citet{K08}; Ni\,\textsc{i} - \citet{K08, WLa}; Cu\,\textsc{i} - \citet{K12}; Zn\,\textsc{i} - \citet{LMW, Wa}; Y\,\textsc{ii} - \citet{BBEHL, K11, PN, HLGBW}; Zr\,\textsc{i} - \citet{BGHL}; Zr\,\textsc{ii} - \citet{LNAJ, CC}; Ba\,\textsc{ii} - \citet{MW}; La\,\textsc{ii} - \citet{LBS}; Ce\,\textsc{ii} - \citet{LSCI}; Nd\,\textsc{ii} - \citet{MC, HLSC}; Eu\,\textsc{ii} - \citet{LWHS}.

\subsection{Line-by-line cleaning}
Lines were accepted or rejected following the rules outlined in \cite{Smiljanic2014}, but with appropriate modifications as this project has results from only two nodes, rather than upward of three nodes for the full {\it Gaia}-ESO WG11 homogenisation. This meant {\it Rules 1, 2,} and {\it 3} of the full {\it Gaia}-ESO WG11 homogenisation process were not applicable here. Also, we did not apply the weights based on the parameters used in \cite{Smiljanic2014} so {\it Rule 7} was also not applicable. The remaining rules (4, 5, 6, 8) were applied as follows: 

\begin{itemize}
   \item {\it Rule 4}: When information of the EWs was available, only lines with 5$\le$EW(m\AA)$\le$120 were used. Exceptions were sodium (5$\le$EW (m\AA)$\le$140) and barium (5$\le$EW(m\AA)$\le$250). This could only be applied for the EW node EPINARBO and no cases outside the limits were found for these stars. \\
   \item {\it Rule 5}: If, for a given species at a given star, abundances from 20 or more different spectral lines were available, we removed the ones that are flagged as blended in the {\it Gaia}-ESO line list (U). This was applied as specified.\\
   \item {\it Rule 6}: If [...] the total number of spectral lines with abundances (for a given species of a given star) is more than 20, a 2$\sigma$ clipping from the mean value was applied. (The total number of lines is counted across all nodes, therefore if eight nodes provide abundances for five lines each, it counts as 40 lines for the clipping). This was adapted to the total lines from the two nodes.   \\
   \item {\it Rule 8}: The median value of multiple lines is adopted as the recommended abundance. This was adapted to be the median of all lines from both nodes.\\
\end{itemize}

The reported spectral lines from each node were `cleaned' so to include only those lines flagged as {\it Y} or {\it U} in the two {\it Gaia}-ESO line list flags {\it gf\_flag} and {\it synflag}, except in the case of {\it Rule 4} above. The definition of the Gaia-ESO line list flags are taken from  \citet{heiterLL} and are as follows:
\begin{enumerate}
    \item {\bf gf\_flag}
    \begin{itemize}
        \item {\it Y}: data which are considered highly accurate, or which were the most accurate ones available for the element under consideration at the time of compilation.
        \item {\it U}: Data for which the quality is not decided
        \item {\it N}: Data which are considered to have low accuracy.
    \end{itemize}    
    \item {\bf synthflag}
    \begin{itemize}
        \item {\it Y}: Line is unblended or only blended with line of same species in both stars (Sun and Arcturus).
        \item {\it U}: Line may be inappropriate in at least one of the stars.
        \item {\it N}: Line is strongly blended with line(s) of different species in both stars.
    \end{itemize}
\end{enumerate}

The reported abundances are thus the median of the remaining line-by-line abundances once the above rules and cleaning were applied. The associated uncertainty was then the MAD converted to a standard deviation by the scale factor of 1.48 ($\sigma$=MAD$\times$1.48).

In the cases of Cr\,\textsc{ii} and Zr\,\textsc{ii} for UVES, and Ca\,\textsc{i}, Cr\,\textsc{i} and Y\,\textsc{ii} for GIRAFFE, only one node provided results for a sufficient number of stars and these were based on one spectral line only. This resulted in an associated uncertainty of zero by the above prescription. In these cases, the uncertainty reported by the node on the spectral line was taken as the uncertainty.

Also, in some cases, the associated uncertainty for a particular star for an element was calculated to be zero while the remaining stars for that element had a non-zero associated uncertainty by the above prescription. In those cases, the median non-zero uncertainty and the MAD converted to sigma on the non-zero uncertainty was calculated from the rest of the sample. A lower threshold was set as the median non-zero uncertainty minus twice the sigma. All associated uncertainties lower than this were set instead to this threshold.

\subsection{Intra-node corrections}
There were seven elements (Mg\,\textsc{i}, Al\,\textsc{i}, Si\,\textsc{i}, Ca\,\textsc{i}, Ti\,\textsc{ii}, Ni\,\textsc{i}, Y\,\textsc{ii},) for which combining the results proved difficult due to systematic offsets between the GIRAFFE and UVES datasets of each node. The seven sets of per instrument and per node results are shown in Fig.~\ref{fig:corabund}.

As there were no stars in common between the GIRAFFE and UVES datasets the way to assess an offset was to look at the overlapping region in \feh{} [-0.8 to -0.4] between the two datasets per node. Two assumptions were made to assess the magnitude and direction of the corrections: 1) Results from a single node should be continuous in \feh{} between the UVES and GIRAFFE results; 2) the Node abundance values around solar metallicity should be similar to the solar abundance.

These allowed the assessment of an offset (difference of the medians across the \feh{} range) and the direction of the correction. For instance, for Mg\,\textsc{i} the EPINARBO GIRAFFE and UVES medians disagree by 0.3 in the overlapping \feh{} region. However, we expect solar Mg\,\textsc{i} at solar \feh{}, hence the correction was applied to the EPINARBO UVES results. This assessment was carried out resulting in the offsets in Table~\ref{tab:abund_offsets} which were then applied at the per-line level. 

\begin{table}
\caption{Node systematic offsets calculated between GIRAFFE and UVES medians across the \feh{} range for seven elements and the node-instrument dataset to which they were applied.}\label{tab:abund_offsets}
\centering
\begin{tabular}{l l l l} 
\hline\hline 
Element & Node & Instrument & Offset \\
\hline 
Mg\,\textsc{i} & EPINARBO & UVES & -0.30 \\
Al\,\textsc{i} & EPINARBO & UVES & -0.37\\
Si\,\textsc{i} & Lumba & GIRAFFE & -0.37 \\
Ca\,\textsc{i} & EPINARBO & GIRAFFE & -0.165 \\
Ti\,\textsc{ii} & EPINARBO & GIRAFFE & -0.215 \\
Ni\,\textsc{i} & Lumba & GIRAFFE & -0.18 \\
Ni\,\textsc{i} & EPINARBO & GIRAFFE & -0.41 \\
Y\,\textsc{ii} & EPINARBO & UVES & +0.185 \\
Y\,\textsc{ii} & Lumba & UVES & +0.41 \\

\hline 
\end{tabular}
\end{table}

\subsection{Comparison of chemical species}
\begin{figure*}
\centering
\includegraphics[width=17.5cm]{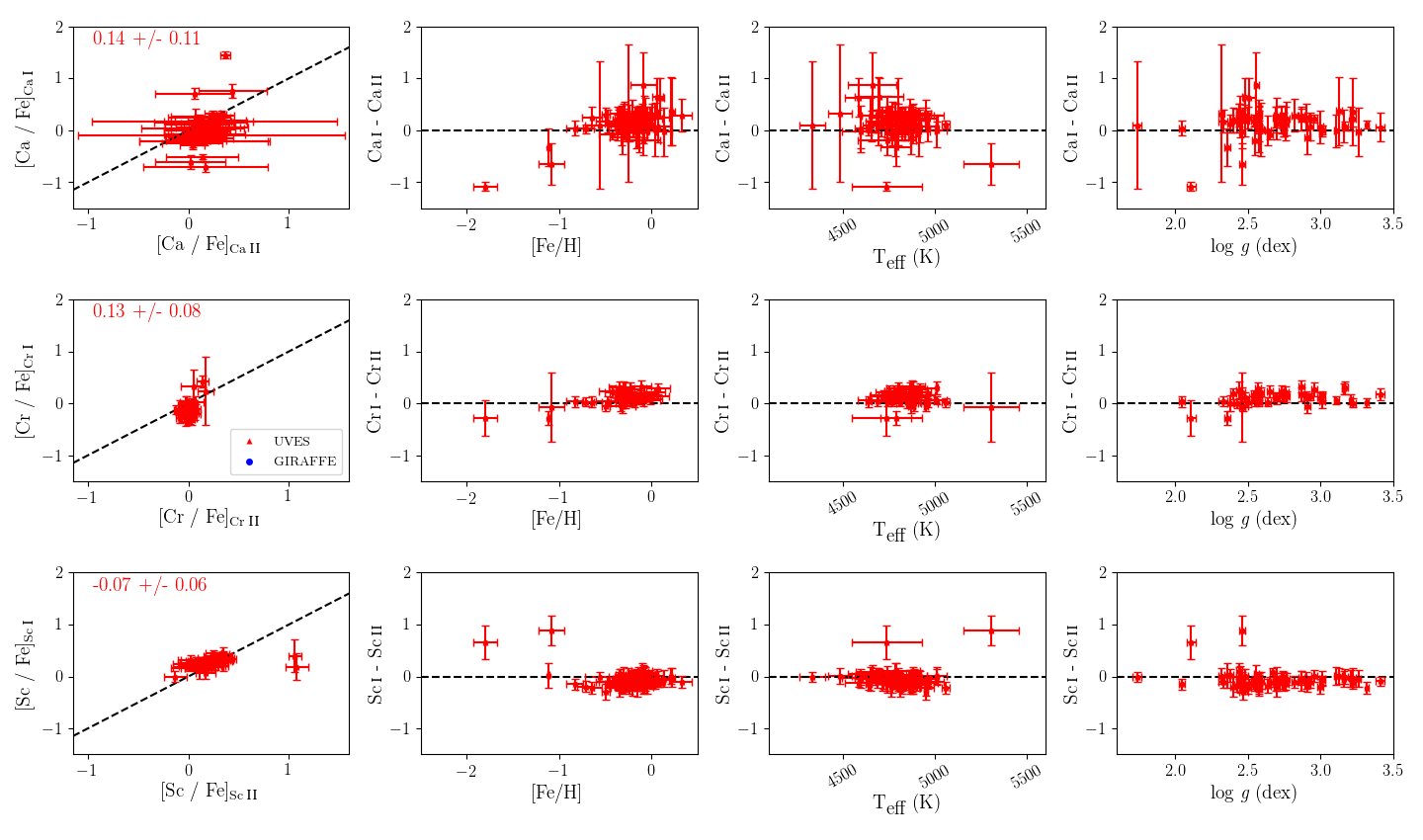}
\includegraphics[width=17.5cm]{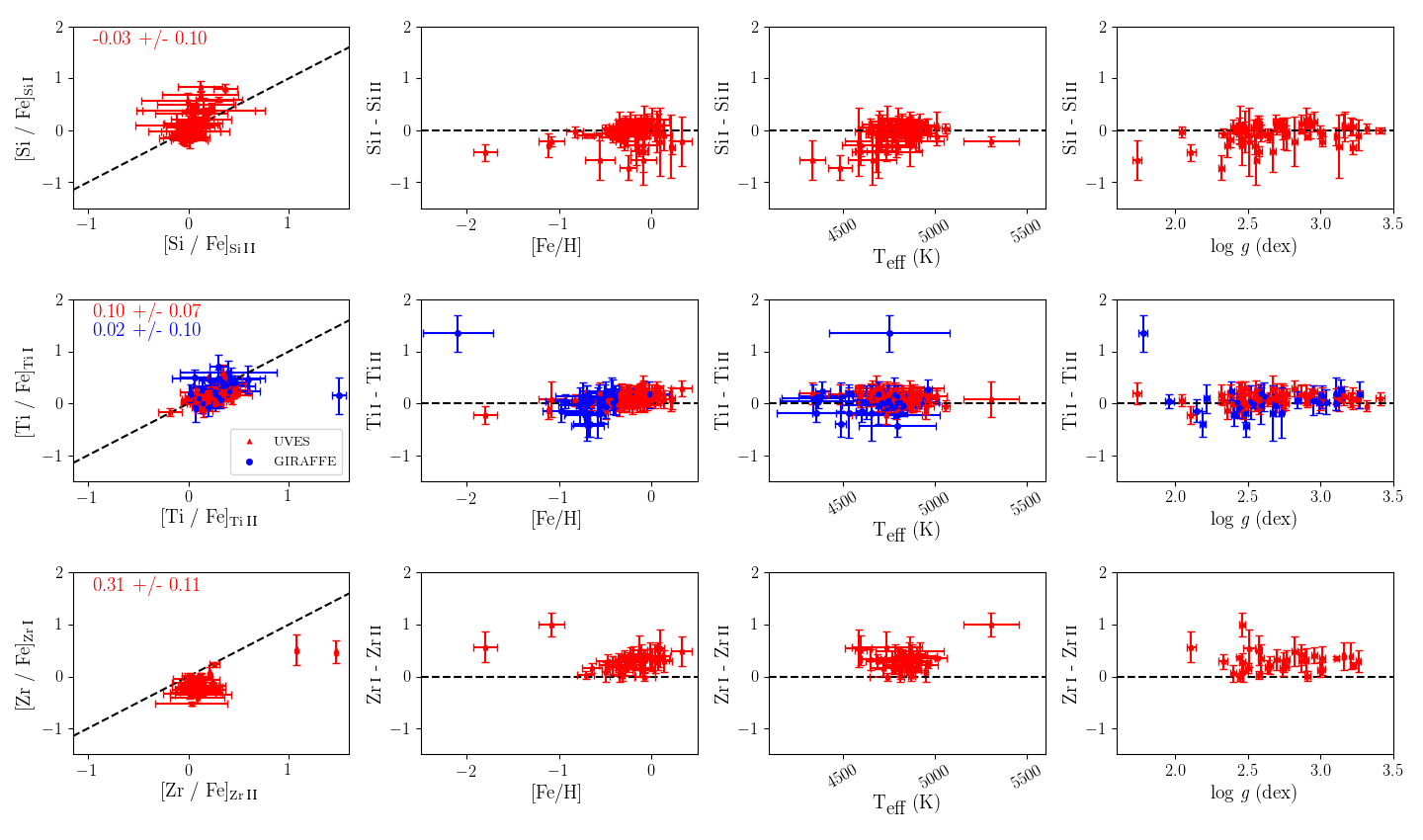}
\caption{Comparing neutral and ionised species of Ca, Cr, Sc, Si, Ti, and Zr as [X\,\textsc{i}/Fe] against [X\,\textsc{ii}/Fe], and X\,\textsc{i}-X\,\textsc{ii} against \feh{}, \teff{}, and \logg{}. Median and MAD of the difference between species of the same element are shown. Red displays the UVES sample, blue displays the GIRAFFE sample. }
\label{fig:species}
\end{figure*}

\begin{figure*}
\centering
\includegraphics[width=18.5cm]{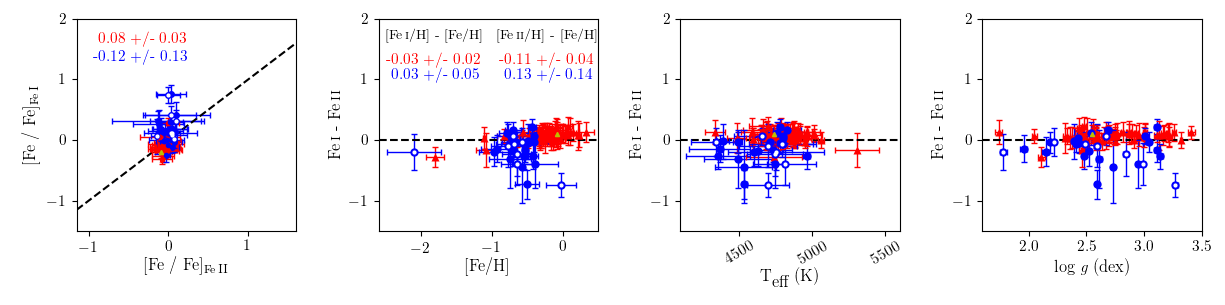}
\caption{Comparing neutral and ionised species of Fe as [Fe\,\textsc{i}/Fe] against [Fe\,\textsc{ii}/Fe], and Fe\,\textsc{i}-Fe\,\textsc{ii} against \feh{}, \teff{}, and \logg{}. Median and MAD of the difference between species of the same element are shown. Red displays the UVES sample, blue displays the GIRAFFE sample. Stars flagged by EPINARBO in the parameter round as having inconsistent Fe\,\textsc{i} and Fe\,\textsc{ii} are indicated as white dots for GIRAFFE and a yellow triangle for UVES.}
\label{fig:species_fe}
\end{figure*}

Seven of the elements considered here have measurements for both the neutral and ionised species: Ca, Cr, Fe, Sc, Si, Ti, and Zr. A comparison between species is shown in Figs.~\ref{fig:species} and \ref{fig:species_fe}. GIRAFFE values (blue points) were only available for Ti and Fe.

In most cases, there is an offset between species on the order of the scatter on the difference. For Zr, for which the median offset is 0.31$\pm$0.11, the largest uncertainties correspond to the stars with the greatest disagreement. Typically for all elements the individual stars with largest difference have a large uncertainty on one of the species measurements. There are no obvious trends with any of the parameters.

Figure~\ref{fig:species_fe} compares Fe\,\textsc{i} with Fe\,\textsc{ii}, then the difference of these against \feh{}, \teff{} and \logg{}. These are the abundances derived for each species of Fe using the final set of stellar parameters. They are distinct from, but based on, the stellar parameter [Fe/H] which is the global metallicity determined in the parameter round. There is excellent agreement between Fe\,\textsc{i} and [Fe/H] for both the UVES (-0.03$\pm$0.02) and GIRAFFE (0.03$\pm$0.05) samples. The agreement for Fe\,\textsc{ii} is less good, showing a larger offset with [Fe/H] for UVES (-0.11$\pm$0.04), as well as a large offset and much larger scatter for GIRAFFE (0.13$\pm$0.14).

In particular, for GIRAFFE, 10 of the 28 stars have a Fe\,\textsc{i} - Fe\,\textsc{ii} > 0.2, which is the median combined error on the abundance differences of the species for the GIRAFFE spectra. Two of these have a difference greater than 0.5.

At the parameter determination stage, EPINARBO flagged stars for which the Fe\,\textsc{i} and Fe\,\textsc{ii} abundances were inconsistent. These are indicated with white dots for GIRAFFE and as a yellow triangle for the single UVES spectrum in Fig.~\ref{fig:species_fe}. While some of these are still outliers, some have been reclaimed as now being consistent between Fe\,\textsc{i} and Fe\,\textsc{ii}.

Of the two with Fe\,\textsc{i} and Fe\,\textsc{ii} abundance differences greater than 0.5, only CNAME 22063424-1530038 was flagged by EPINARBO in the parameter round as having inconsistent Fe\,\textsc{i} and Fe\,\textsc{ii} abundances. There was also a difference in Spec2 \teff{} between the nodes of 504~K. Consequently only the Lumba parameters were used for the final parameters (see Table~\ref{tab:fin_stellar_params}). This star has an outlying rotational velocity which is discussed further in Sect.~\ref{sec:induce_vrot}.

CNAME 22071427-1431390, also with Fe\,\textsc{i} - Fe\,\textsc{ii} > 0.5, was not flagged as an outlier during the parameter stage, although there is a difference in the Spec2 \feh{} values between Lumba and EPINARBO of -0.24. It has, however, been flagged as a potential binary (see Sect.~\ref{sect:binaries}). Two other spectra with Fe\,\textsc{i} - Fe\,\textsc{ii} > 0.5 were also flagged as potential binaries: CNAMEs 22003290-0808595, 22031541-0753433.

To further understand the possible source of inconsistencies for the remaining seven GIRAFFE stars, for GIRAFFE, the Fe\,\textsc{ii} value is based on a maximum of two spectral lines and only comes from the EPINARBO analysis, compared to at least nine lines for the Fe\,\textsc{i} value from both EPINARBO and Lumba (see Table~\ref{tab:listelements}). Additionally, the Fe\,\textsc{ii} lines are also typically weaker and thus more difficult to measure in medium- compared to high-resolution spectra, resulting in less certain values as indicated by the larger uncertainty bars.

For UVES, 3 of the 62 stars have a Fe\,\textsc{i} - Fe\,\textsc{ii} > 0.16, which is the median combined error on the abundance differences of the species for the UVES spectra, although none of these have a difference greater than 0.3. None of these three (CNAMEs 22065112-1504580, 22195215-1234594, 22021848-1139147) were identified as an outlier in the parameter homogenisation. CNAME 22065112-1504580 has the largest difference between Fe\,\textsc{i} and Fe\,\textsc{ii} of -0.28 and it is the only one of these three to be flagged as a potential binary (see Sect.~\ref{sect:binaries}).

We note that the ionisation equilibrium (consistent abundance values between neutral and ionised species) may not have been achieved in these cases because the $\log g$ value was constrained by non-spectroscopic data, which may reflect shortcomings of the atmospheric models used in the spectroscopic analysis, such as the LTE and 1D prescriptions.

\subsection{K2@{\it Gaia}-ESO final chemical abundances}
The final chemical abundances for the K2@Gaia-ESO stars with the above corrections applied are shown in full in Fig.~\ref{fig:finabund}. The full catalogue table is available online but the columns are provided in Table~\ref{tab:onelinetab_cols}. The UVES (high-resolution) and GIRAFFE (medium-resolution) measurements are shown separately. In general the two resolutions track each other for each element distribution, with the improvements due to the corrections applied above. Overall, there appears to be greater scatter in the GIRAFFE datasets, which is not unexpected, particularly for those elements where the measured spectral lines are weak or more blended at lower resolutions \citep[see figures in e.g.][]{Jofre18}.

\begin{figure*}
\centering
\includegraphics[width=18cm]{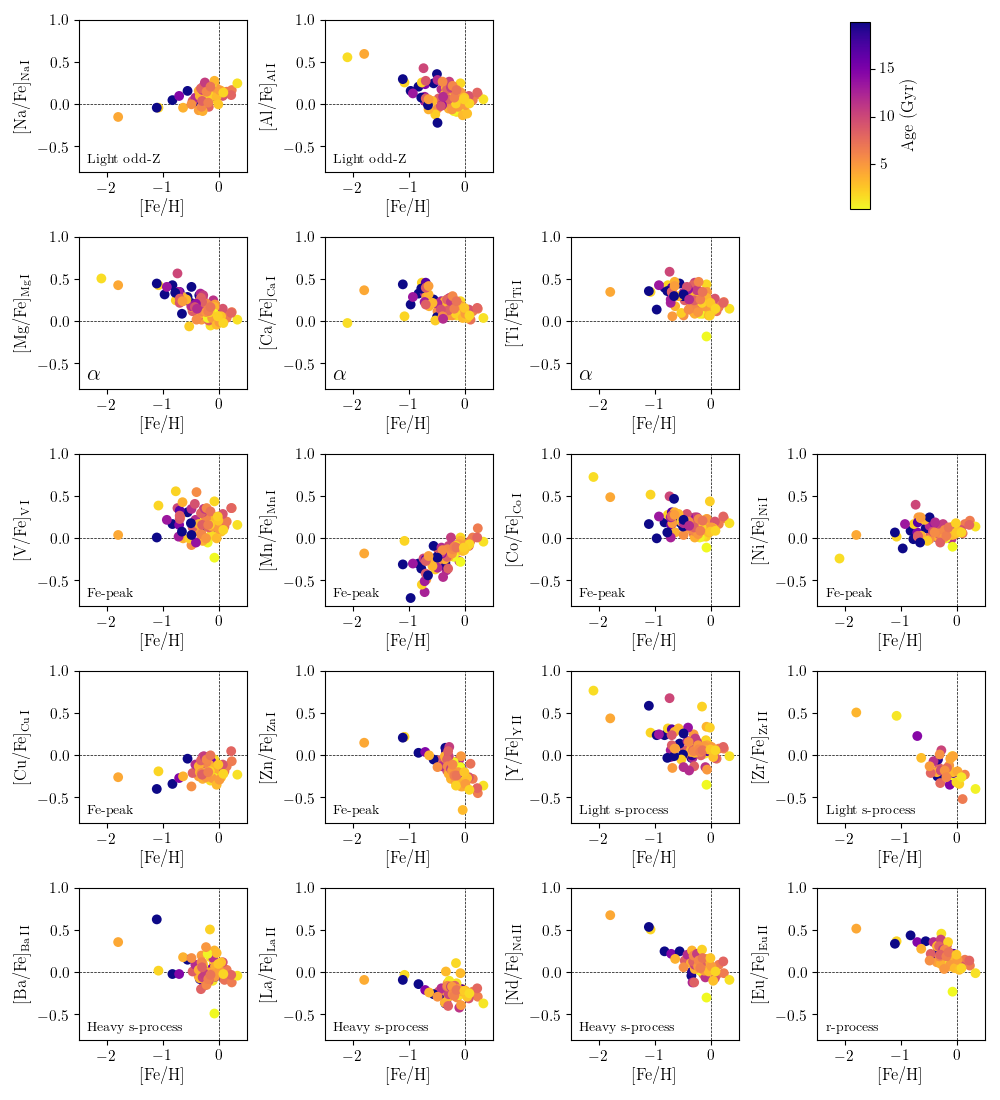}
\caption{Chemical abundances of [X/Fe] against \feh{} for the K2@{\it Gaia}-ESO stars with a colourmap of Age. \feh{}=0 and [X/Fe]=0 are indicated as dashed lines.}
\label{fig:agefinabund}
\end{figure*}

The final stage of this work is focussed on the derivation of ages and masses within the greater K2 Galactic Caps Project (see Paper~I). Specifically for this paper, the ages of the K2@{\it Gaia}-ESO sample were derived, shown as a colour map on the chemical distribution of key elements in Fig.~\ref{fig:agefinabund}.

This set of elements is arranged in five key nucleosynthetic channels: Light odd-Z, $\alpha$, Fe-Peak, s-process and r-process. Each of these channels reflect different nucleosynthetic origins in stellar evolution. Elements created in the same channel show similar behaviour in chemical distribution morphology. These are interpreted in consideration of the recent release of the GALAH DR2 abundances presented in \citep{buder2019} which used a pipeline which was developed based on the experience with the Lumba pipelines. We note that this is a qualitative comparison only. On the one hand, there are no stars in common between this sample and GALAH DR2. In fact, Gaia-ESO and GALAH only have a small overlap in observed targets (the Gaia benchmark stars, some targets in M67 and the CoRoT fields). On the other hand, our methodology differs from the standard Gaia-ESO parameters so direct comparison between surveys is not the aim here.

Considering the two Light odd-Z elements,  Na\,\textsc{i} shows the expected trend of depleted Na at low metallicity increasing to $\sim$0.15 at super-solar metallicity. Its distribution is similar to Ni\,\textsc{i} and Cu\,\textsc{i} as expected from previous studies \citep{buder2019}.
On the other hand, Al\,\textsc{i} shows enhanced Al at low metallicity, although no non-LTE corrections have been applied contrary to \cite{buder2019}. Al behaves similarly to Mg and the other $\alpha$ elements, also noted in \cite{buder2019}.
The $\alpha$ elements presented here (Mg\,\textsc{i}, Ca\,\textsc{i}, Ti\,\textsc{i}\footnote{Ti behaves observationally like the rest of the $\alpha$ elements}) all bear the same morphology of showing enhancement at low metallicity decreasing to solar at solar metallicity, which is the typical chemical distribution of $\alpha$ elements  \citep{buder2019}.

The Fe-peak elements (V\,\textsc{i}, Mn\,\textsc{i}, Co\,\textsc{i}, Ni\,\textsc{i}, Cu\,\textsc{i}) show at least four types of morphology. V\,\textsc{i} is unique in that it stays generally constant with a large scatter over the presented metallicity range, which is consistent with other studies. Mn\,\textsc{i} also is consistent with other studies showing a clear trend of increasing Mn\,\textsc{i} with metallicity. Co\,\textsc{i}, Ni\,\textsc{i} and Cu\,\textsc{i} can be grouped with similar if not the same morphology of slightly increasing at the metal-poor end, then slightly decreasing then slightly increasing again at solar. This is also not dissimilar to Na\,\textsc{i} as noted above. The overall increasing trend with metallicity follows that presented in other studies \citep{buder2019} although the behaviour around solar begs further investigation.

The s-process is separated into the light (Y\,\textsc{ii} and Zr\,\textsc{ii}) and Heavy (Ba\,\textsc{ii}, La\,\textsc{ii}, Nd\,\textsc{ii}) s-process peaks. Y\,\textsc{ii} shows a large scatter with there being enhancements and depletions for both old and young stars. This was similarly found in \cite{buder2019}. There are fewer measurements for Zr\,\textsc{ii}, particularly at the metal-poor end, but a similarly large spread is indicated.

For the heavy s-process, there are only few measurements at the metal-poor end but these mostly indicate enhanced levels for metal-poor stars for Ba\,\textsc{ii}, La\,\textsc{ii} and Na\,\textsc{ii}. Despite\ showing a larger scatter for Ba\,\textsc{ii}, all are fairly constant around solar. Finally, the single r-process element, Eu\,\textsc{ii}, shows enhancements at low metallicity, decreasing to a constant solar values at solar and super-solar metallicities, not unlike the $\alpha$ distributions.
Although the sample is small compared to other samples where spectroscopy and seismology is available \citep[e.g.][]{apokasc2}, even for this sample of 90 stars, the morphology of the different nucleosynthetic channels can be seen. 

\begin{figure*}
\centering
\includegraphics[width=18cm]{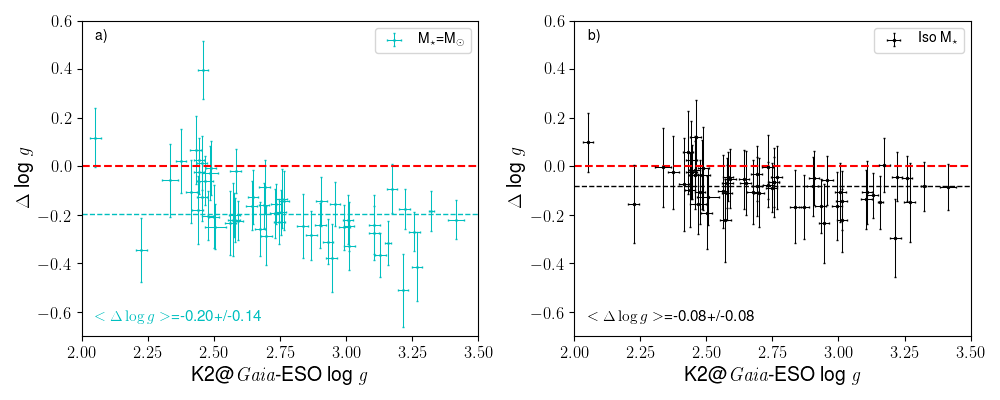}
\caption{Comparison of \logg{} determined using {\it Gaia} parallax with the final K2@{\it Gaia}-ESO seismic \logg{}. Cyan points are derived from an initial mass equal to solar, black points are derived from masses based on isochrones. The dashed red line shows the zero difference and the dotted black and cyan lines shows the mean of the differences respectively.}
\label{fig:gaia_k2_parallax}
\end{figure*}

The key outliers in this sample are the two most metal-poor stars which appear to be quite young ($\sim$2~Gyr). This is discussed further in Sect.~\ref{sect:binaries}.

\section{K2@{\it Gaia}-ESO@{\it Gaia}: Parallaxes}
\label{sect:parallaxes}
The iterative determination of the seismic \logg{} and spectroscopic \teff{} and \feh{} was completed prior to the release of {\it Gaia} DR2 in April 2018. However, the parallaxes from {\it Gaia} DR2 provide a useful check on the robustness of the final K2@{\it Gaia}-ESO surface gravities. 

The \logg{} based on the {\it Gaia} parallax was calculated using the following additional inputs: the final K2@{\it Gaia}-ESO \teff{} and \feh{}; the reddening, E(B-V), assigned to each target as part of the {\it Gaia}-ESO dataset was taken from \cite{Schlegel1998}; and the 2MASS K band photometry \citep{2MASS}. Of the 90 stars, only 58 are compared here due to the {\it Gaia} parallax either not being available or not good enough (parallax uncertainty was greater than 15\%).

An iterative procedure using isochrones to determine the optimal stellar mass and surface gravity based on the {\it Gaia} parallaxes was carried out (\logg{}$_{\textrm{iso}}$). Further details on the calculation are provided in Appendix~\ref{app:gaialogg}. For comparison, the surface gravity based on the {\it Gaia} parallax assuming only solar mass was also calculated (\logg{}$_{\textrm{M}_{\odot}}$).

Figure~\ref{fig:gaia_k2_parallax}a compares \logg{}$_{\textrm{M}_{\odot}}$ with the final K2@{\it Gaia}-ESO \logg{} and Fig.~\ref{fig:gaia_k2_parallax}b compares \logg{}$_{\textrm{iso}}$ with the final K2@{\it Gaia}-ESO \logg{}. The mean offset for each are $\Delta$\logg{}$_{\textrm{M}_{\odot}-\textrm{K2}}$ -0.20$\pm$0.14 and $\Delta$\logg{}$_{\textrm{iso}-\textrm{K2}}$ -0.08$\pm$0.08 respectively. Overall \logg{}$_{\textrm{iso}}$ are in better agreement with the K2@{\it Gaia}-ESO \logg{} than \logg{}$_{\textrm{M}_{\odot}}$. The typical reported \logg{} uncertainties of K2 are ~0.02 and for the {\it Gaia} analysis are ~0.13. Spectroscopic \logg{} typical uncertainties are 0.10 to 0.25 for the {\it Gaia}-ESO high-resolution UVES spectra \citep{Smiljanic2014} and 0.15 to 0.40 for the {\it Gaia}-ESO medium-resolution GIRAFFE spectra \citep{worleywg10}.

It is known that the parallaxes provided by {\it Gaia} DR2 are affected by systematics \citep{gaiadr2lindegren,zinn2019}.
A comparison of {\it Gaia} parallaxes and K2 seismic parallaxes for stars in the C3 field in \cite{Khan} found an offset ranging between -45 to -55 $\mu$as. The authors found that the correction in the parallax between {\it Gaia} and seismic values seems to depend, to varying degrees, on the position on the sky, the magnitude, and, potentially, the colour. Also, specifically for the K2 C3 and C6 fields, the correction seems to be smaller than in the Kepler field. Following the recommendation in \cite{Khan}, no ad hoc correction was attempted here but the possible systematics are noted as follows.

Using the equations in Appendix~\ref{app:gaialogg}, the possible range in offsets in parallax corresponds to a range in offsets in \logg{}$_{\textrm{{\it Gaia}-K2}}$ of -0.05 to -0.07. This is in the same direction and of similar magnitude to that measured here, showing the offset between parallaxes from {\it Gaia} DR2 and asteroseismology found in this study is in agreement with the previous studies \citep{zinn2019,Khan}.

The main outlier is CNAME 22235003-1422417 with $\Delta$\logg{}$_{\textrm{{\it Gaia}-K2}}=-0.30\pm0.16$. This star was observed with UVES and had not been flagged as particularly peculiar otherwise. The seismic \logg{} values for each node were in good agreement which is not unexpected as shown above. The difference between the node \teff{} values is LM-EP=-130~K and \feh{} values is LM-EP=-0.39. These are within the acceptable limits imposed in this study for classifying outliers. Three other stars have $|\Delta$\logg{}$_{\textrm{{\it Gaia}-K2}}|>0.20$.

\section{K2@{\it Gaia}-ESO@{\it Gaia}: binary stars}
\label{sect:binaries}


Both the spectra of {\it Gaia}-ESO and the astrometric measurements of {\it Gaia} offer important information about a star's movement. These can be combined to study potential binary stars in the sample.  This information is given in Table~\ref{6xGaia DR2}.

We compared the radial velocities (RV) of {\it Gaia}-ESO with those from {\it Gaia} \citep{gaiarvsprocessdr2,gaiarvsvaliddr2} from the {\it Gaia} Radial Velocity Spectrometer \cite[RVS:][]{gaiarvs}. We generally find a very good agreement, with a zero point offset of $-0.44$~km/s for UVES and of $-0.95$~km/s for GIRAFFE, based on which we infer that {\it Gaia}-ESO provides systematically lower velocities.  The offset is determined from the mean of the RV differences between {\it Gaia} and {\it Gaia}-ESO. We note the relatively large offset for the GIRAFFE case is based on very few measurements. Indeed, the faint magnitudes of the GIRAFFE sample make the overlap with {\it Gaia} RVS very small. 

\begin{figure}
\centering
\includegraphics[width=10cm]{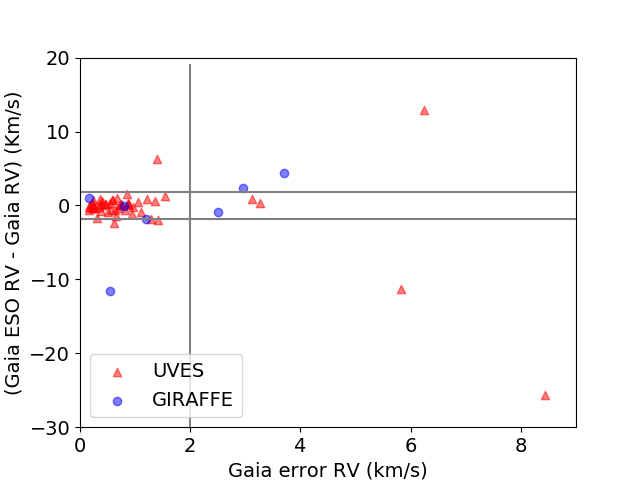}
\caption{Comparison of radial velocities of {\it Gaia}-DR2 and {\it Gaia}-ESO as a function of the uncertainty reported by {\it Gaia}. Stars that have larger uncertainties than the vertical line at 2~km/s, or a larger RV difference than the horizontal lines at 2~km/s, are classified as binaries in this work. }
\label{fig:rvs}
\end{figure}

Individual comparisons are shown in Fig.~\ref{fig:rvs}, in which we plot the difference of the reported velocities from {\it Gaia}-ESO and {\it Gaia} as a function of the {\it Gaia} uncertainty for all stars that are contained in both datasets. While the majority of the stars have uncertainties in {\it Gaia} that are below 1 km/s, there are a few stars that are more uncertain. There is, however, no relation between {\it Gaia} RV uncertainty and apparent magnitude. The values used for the figure can be found in Table~\ref{6xGaia DR2}.

In addition, by taking a conservative cut of (1) an uncertainty reported by {\it Gaia} larger than 1.8~km/s or (2) a difference in radial velocities between {\it Gaia} and Gaia-ESO larger than 1.8~km/s, we can select potential binaries. In Fig.~\ref{fig:rvs}, lines marking an uncertainty or difference of 1.8~km/s are shown vertically and horizontally. It is clear that some stars do not agree that well, falling outside the box marked by the lines within 1.8 km/s.  These measurements might be attributed to binary stars. We note that the reported {\it Gaia} RV corresponds to the median of each of various RV measurements at different epochs and its uncertainty may be attributed as the variability in RV for each target.  {The number of transits varies a lot between each {\it Gaia} target, as}  indicated in Table~\ref{6xGaia DR2}. A large uncertainty might be related to a large variability, which can be an indication of binary systems \citep[see e.g. recent paper on comparisons of RV of Gaia and RAVE by][]{birko19}. 

We selected our binaries as those with differences in RV of at least 1.8 km/s to account for the intrinsic variation in RV due to jitter, time-variable winds, spot visibility, and so forth, which can have a notable effect in evolved giant stars \citep[see e.g. ][for a reference about jitter on metal-poor giants]{carney2003}, and for differences of zero point between Gaia-ESO and Gaia, which in our case is for both datasets below 1~km/s. 
This gives a total of 14 potential binaries in our sample, corresponding to 15\%. This value agrees with the estimate of 17\% for metal-poor giants found by \cite{carney2003} but the comparison should be taken as a reference only since both samples are selected entirely differently. Indeed, the fact that we have been able to provide parameters for the stars implies that no peculiarity in the spectra has been found. It is important to remark that this comparison is also not fully representative as {\it Gaia}-DR2 delivered RVs only for stars that are not double-lined spectroscopic binaries. {\it Gaia}-ESO, on the other hand, determines RVs for every star, regardless of their binary status. Indeed, a special working group within {\it Gaia}-ESO (WG14) analyses double-line spectroscopic binaries and peculiar stars, flagging every target prior to the spectral analysis \citep{merle2017, Merle2020}. None of the stars in this sample have been flagged as a double-line spectroscopic binary or as a peculiar star by WG14. 

It is worth commenting here that we aimed to identify and investigate the potential binaries that this sample might contain. Yet, variable RV is not the only indication that the star may be a binary. If stars have relatively small RV variation and are binaries at the same time, the companion is likely to be less massive and, so, it exhibits larger RV excursions. As a result, the lines from the companion fall in the wings of the lines of the primary star so that the spectrum appears to be that of a fast rotator, even though in reality it is a set of blended lines. This is seen also in RAVE where the spectrum of a cool dwarf appeared to be a fast rotator, but was ultimately found to be a binary. A similar effect might exist for giants.

\begin{figure}
\centering
\includegraphics[width=10cm]{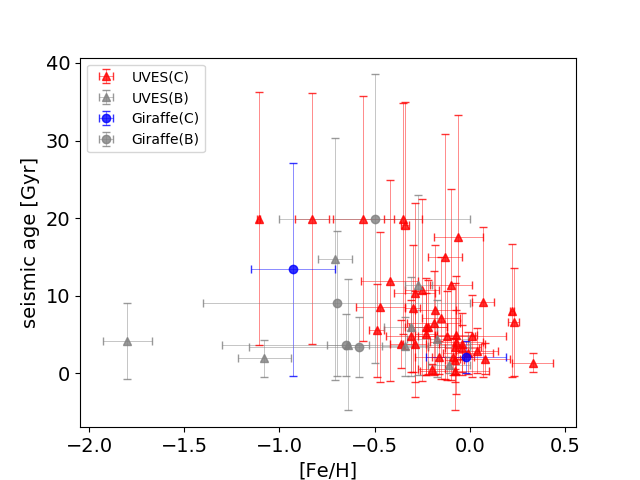}
\caption{Age-metallicity relation of the stars in our sample. Grey indicates the stars that are binaries (B) and coloured symbols indicate stars that are constant (C) according to our classification (see text). Error bars in age represent the mean between upper and lower age uncertainty estimate (see Paper I).}
\label{fig:age_feh}
\end{figure}

In Fig.~\ref{fig:age_feh}, we plot the ages and metallicities of the stars, coloured according to the respective instrument and indicating in grey the stars that are binaries according to our classification. In general, the expected age-metallicity relation is found, in which metal-poor stars are old and metal-rich stars, can have a wide range of ages. We note that the oldest stars in our sample show ages older than the Universe. As discussed in Paper I, this reflects that the prior used for the age determination was set to a maximum of 20 Gyr. When uncertainties of the results are taken into account, the oldest stars in our sample have ages within the age of the Universe. Here we plot the absolute values with the intention to study their distribution but full details can be found in Paper I. 

We see that two metal-poor stars show notably young ages. For one of them, we have further RV information from Gaia and we note that this star is classified as a possible binary. This star, along with other binary candidates that stand out, are discussed further below.

\subsection{EPIC 206101493: potentially an evolved blue straggler}

\begin{figure}
\centering
\includegraphics[width=10cm]{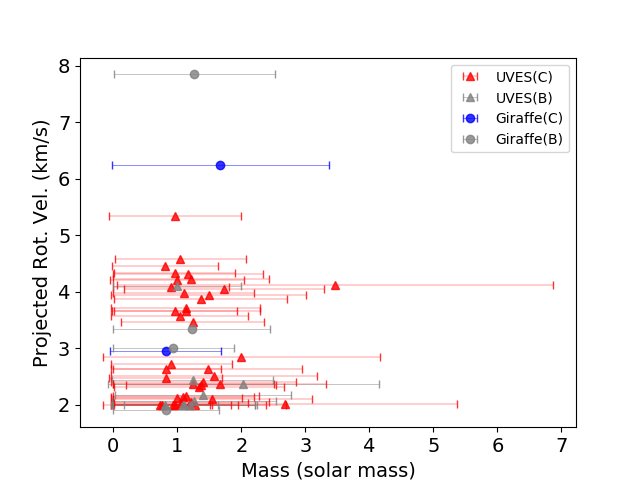}
\caption{Mass and projected rotational velocity ($v \sin i$) for those stars which have Gaia RV measurements. Error bars in mass represent the mean between upper and lower mass uncertainty estimate (see Paper I).  }
\label{fig:vrot}
\end{figure}

As in several other recent works, some metal-poor (and alpha-enhanced) stars show unexpectedly young ages. The case of EPIC 206101493 is one example found in this K2@{\it Gaia}-ESO sample. Its CNAME is 22000793-1203412 and has been observed with UVES.  
 
Such apparently young stars have been reported by \cite{Chiappini15}, and were called 'young alpha-rich stars' (YAR). The observation that their high masses might imply they are young \citep{Martig15} and their chemical composition is more consistent with that of thick disk stars \citep{Matsuno18} challenges our belief that the thick disk formed a long time ago and very rapidly.  It is thus very important to first test if they are (or were) binaries, as this determines whether we can or can not apply stellar models of isolated objects to constrain the age of YAR stars. This is discussed in \cite{Jofre16} with the help of RV monitoring of the YAR stars listed by \cite{Martig15}, and then studied in \cite{Izzard18} with the help of stellar synthesis models which include the interaction of binary stars.  It is thus interesting to see that in our K2@{\it Gaia}-ESO sample, EPIC 206101493, as another identified YAR star, is flagged as a binary due to its RV inconsistencies between {\it Gaia}-ESO and {\it Gaia}. 

EPIC 206101493 has a UVES spectrum but since it is one of the faintest stars in our sample, its spectrum has a very low S/N ($\sim 29$), this does not allow us to perform a very detailed chemical analysis to investigate its properties further, like \cite{Matsuno18} did for a sample of these stars. In any case, at a first glance, its parameters are well-determined and its abundances are consistent with a thick disk star. It presents no evident enhancements of s-process or Li. This means that EPIC 206101493 is as typical a thick disk star as the other YAR stars studied in \cite{Matsuno18}, and its high mass agrees with what has been postulated by \cite{Jofre16}, namely, that it is likely a product of mass transfer.

\subsection{EPIC 206322094 and EPIC 205977363: induced rotation}\label{sec:induce_vrot}
In Fig.~\ref{fig:vrot}, we show the mass and the projected rotational velocity ($v\sin i$) for our stars, following the same symbols as in previous figures.  The projected rotational velocity is determined by {\it Gaia}-ESO alongside the radial velocity through the cross-correlation with templates. This is similar to the procedure reported by \cite{Carney08} for their study of line broadening and rotation of metal-poor giants. 

We can see that the giants in our sample have typical $v\sin i$  of 2-4~km/s, regardless of their binary status or stellar mass. There are however two stars that stand out, with $v \sin i > 6$~km/s (EPIC 206322094 $v \sin i = 7.85$~km/s; EPIC 205977363 $v \sin i = 6.25$~km/s). We note there are other stars in our sample with even higher $v \sin i$ but they are excluded from this discussion since they do not have RVs from Gaia DR2. While this value is higher than the rest of the stars in the sample, an abundance analysis is still possible and could be done safely. 

Giants with these values of rotation have been classified as `anomalous rotators' by \cite{tayar15} in the APOKASC sample. They are attributed to be the cause of a recent interaction. As extensively discussed by \cite{Carney08}, the higher broadening of these stars is not likely to be due to macroturbulence. There is a trend for macroturbulence with stellar parameters which is not followed for the outlier cases. This suggests that something else might be causing the broadening. \cite{carney2003} offer other interesting explanations for outstanding line broadening which are worthy of note. 

EPIC 206322094 (CNAME 22031541-0753433) is marked as a binary, and its line broadening might be reflecting an induced rotation due to tidal interactions. {\it Gaia} reports an uncertainty in RV of 2.5~km/s. If the star is tidally interacting with a companion, then its orbit must be circular, which can be tested with individual RV measurements. Unfortunately, this information is still not available from {\it Gaia}, but a follow-up campaign on this star is ongoing and will be reported on in a separate paper. 

The other star with outlying rotational velocity is EPIC 205977363 (CNAME 22063424-1530038), however, this star does not show indication of having a binary companion from its radial velocity behavior. The star could be the result of a merger \citep{tayar15}. Another interesting possibility for its increase in $v_{\mathrm{rot}}$ is the capture of a giant planet of few Jupiter masses due to which the star grew in size and became a red giant. It has been shown in the literature that this process can spin up giants a few km/s depending on the planet mass and the stellar radius \citep{Privitera2016}. 

\cite{siess&livio99} have  studied this scenario extensively, discussing possible effects that can be observed. One effect of planet capture is mass loss, which can be detected by the presence of an emission feature or asymmetry in the H$\alpha$ line. Alternatively, recent discussion in the literature connects Li-rich giants to planet engulfment, where the Li abundance now unexpectedly present in the star was initially material from the planet \citep[see e.g. discussions in ][using {\it Gaia} ESO data, and references therein]{casey16}. Although it is not entirely clear  that engulfment might explain all Li-rich stars \citep[see e.g. ][]{aguilera16}.  Since this star is observed with the GIRAFFE setup, we do not have information about the Li abundance or the H$\alpha$ profile unless a spectrum covering this wavelength domain is undertaken in the future.

\subsection{EPIC 206070270 and its unseen companion}
In our sample, we have one star (EPIC 206070270 - CNAME 22182719-1252466) with a notably large uncertainty in RV as determined by {\it Gaia} (above 8 km/s). The radial velocities measured by {\it Gaia}-ESO and {\it Gaia} differ by 25 km/s, which is consistent with the large RV uncertainty. While this is probably due to it being a binary system, without single RV measurements it is difficult to know for sure. However it is possible that this star is similar to the binaries identified in APOGEE by \cite{price-whelan18}.  In their catalogue, they use the RV variation (and single epoch RV visit) of APOGEE giants to find the properties of possible binaries. The spectrum is further used to determine the stellar parameters and hence the mass of the giant. The stellar masses they determined cover a range that encompasses the mass of EPIC 206070270 as determined by seismology. Furthermore, they show examples of giants with high variations in RV, in some cases reaching the 25~km/s we find here. This value is also a typical value for giants found by the study of mutiplicity of giants in APOGEE by \cite{badenes18}. 
If we consider that this difference is the maximum amplitude of the RV variation, and that the giant has a mass of $1.2~\mathrm{M}\odot$ (see Paper I), we can use the information in Table 5 of \cite{price-whelan18} to estimate that the companion of EPIC 206070270 must be a star with approximately half a solar mass. 

The fact that we can constrain the mass and age of the primary with asteroseismology makes this system a potentially interesting target for follow-up studies of RV to constrain the orbits and masses of the secondary. This campaign is taking place already with the HERMES telescope in La Palma and the results will be presented elsewhere once enough RV measurements and a sufficient time span has been achieved to robustly determine the parameters of the binary system.


\section{Discussion and conclusion}

This work presents a detailed study to determine the stellar parameters of 90 red giant stars by combining results from spectroscopy and asteroseismology. The spectroscopy comes from data taken as part of the {\it Gaia}-ESO Survey in a project designed to expand the available set of calibration samples. The asteroseismology comes from data taken by K2 in the C3 field.  The parameters could therefore be determined with higher accuracy due to the iterative procedure of deriving \teff{} from the spectra to be used as input for deriving \logg{} from oscillation frequencies which is then used to re-derive \teff{}. A substantial change in \teff{} results in just a small change in the seismic \logg{} and so both methods are required to ensure the required accuracy in parameters.

Understanding the origin of the Galactic disks is based on a fundamental question and the answer  is sought in the exploitation of stellar surveys. However, there are underlying challenges due to the inaccuracy of the stellar parameters and abundances derived for metal-poor and ancient stars. Having as many reference stars as possible in this regime is key to finding the answer.

The C3 field was chosen as it points towards the South Galactic Pole, thus favouring the inclusion of thick disk and metal-poor stars. These stars are under-represented in other seismic fields, notably the {\it Kepler} field (see \cite{rendlek2ges} for the space distribution of the stars of K2C3 and {\it Kepler}). In addition, a study of chemical abundance evolution and binary detection was performed given the additional information pertaining to the masses and ages from K2, as well as the information of {\it Gaia} DR2. 

The spectroscopic analysis was based on the combination of two distinct methods, both of which are part of the analysis procedure of the {\it Gaia}-ESO data. One method was based on spectrum synthesis and the other one was based on equivalent widths. Together, they allowed us to assess systematic uncertainties and reliability of results, thus providing a catalogue of spectral parameters and abundances that is accurate and not biased towards a given methodology. However, based on the intra-node abundance assessment and the corrections that were needed for seven of the elements, some possible systematic offsets in individual element abundances may remain. 

The combination of seismology and spectroscopy allows us to analyse trends of abundances as a function of age, finding, as expected, that more metal-poor and $\alpha$-capture element-enhanced stars are older. The addition of the {\it Gaia} DR2 data also allows us to identify possible binary stars: the ones that are suspiciously more massive for their metal-content, those rotating unusually faster, or those having differences in radial velocities that are larger than expected. 

Catalogues of parameters and abundances for which both seismology and spectroscopy are available are few. Pioneering works in this direction are the APOKASC sample \citep{apokasc, apokasc2} and the CoRoGEE sample \citep{Anders2017}. \cite{Hawkins16a} published a catalogue with abundances for this sample considering the infrared APOGEE spectra based on temperatures  determined photometrically  and surface gravities from scaling relations of {\it Kepler}. 

Another group at the forefront of this work is led by \cite{valentini17}, who published a study similar to the one presented here (i.e. iterating between spectroscopy and seismology to reach to consistent parameters). They analysed 87 giants observed by RAVE. The spectra are of intermediate resolution (R~7000) and cover a short wavelength range around the Ca\,\textsc{ii} triplet, equivalent to  {\it Gaia} RVS. In \cite{valentini17} it is possible to see how the overall stellar parameters improve when seismology is available. The group further performed a follow-up, detailed,  high-resolution study of four metal-poor stars of their sample, which is presented in \cite{valentini18}.  The full RAVE catalogue will soon become public \citep{steinmetzrave}. Targets with seismology have proven to be fundamental for the calibration of surface gravities of RAVE  \citep{kunder17}.  Similarly, {\it Gaia}-ESO is in the process of using CoRoT data as part of the calibration plan \citep{pancino17}, with the intention to publish the CoRoT@{\it Gaia}-ESO catalogue \citep{masseeroncorot}.  In  \cite{valentini16}, the UVES and GIRAFFE stars were analysed following the same procedure as for the RAVE study. 

Recently, \cite{Sharma19} presented a study of the age and metallicity gradient of a large sample of K2 stars observed in GALAH. The study includes the same fields we analyse here, but GALAH spectra are currently not public to the entire community. The parameters and abundances are part of GALAH DR2 \citep{buder18}, although they were not derived using the seismic parameters as a prior.  

In addition, \cite{nissen17} published a catalogue of parameters and abundances derived self consistently with seismology using high-resolution optical spectra of HARPS-N for a sample of solar twins in {\it Kepler}. Finally, \cite{morel14} presents the study of high resolution optical spectra of 20 red giants that were targeted by CoRoT. 

Our work is a novel contribution as the largest high-resolution optical catalogue with parameters and abundances derived homogeneously. We have taken advantage of forefront spectroscopic and astereoseismic methodologies, along with the sample targets of the outer regions of the disk to capture sought-after old and metal-poor stars. Our catalogue of parameters and abundances is available online, including the results at a line-by-line and method-by-method level to ensure reproducibility (see Tables~\ref{tab:onelinetab_cols} and \ref{tab:lbltab_cols} for list of columns). The spectra are public via the ESO archives. In addition, the chosen field of K2 was selected so that our catalogue would include more thick disk stars than other samples for which spectroscopy and astereoseismology are available. 

While the interior of stars gives us clues as to their evolutionary phase (based on mass and age), the exterior of stars tell us about their formation site (chemistry), and stellar motion (radial velocities). Individually, these latter two types of data can provide the means to calibrate the analysis of large samples of stars, however, the combination of both provides an even more powerful basis upon which to make a more effective and robust analysis that is key to constraining models of galactic formation and evolution.

\label{sect:conclusion}

\bibliographystyle{aa}

\section{Acknowledgements}
The authors thank the International Space Science Institute in Bern (ISSI) and Beijing (ISSI-BJ) for supporting and hosting the meetings of the International Team on “AsteroStep” and "Libraries of 2020", respectively, during which the discussions leading and contributing to this publication were initiated and held.

PJ acknowledges support of FONDECYT Iniciaci\'on Grant Number 11170174. RS acknowledges support from the National Science Centre (2014/15/B/ST9/03981). A.G., A.K. and U.H. acknowledge support from the Swedish National Space Agency (SNSA/Rymdstyrelsen). T.B. was funded by the project grant “The New Milky Way” from the Knut and Alice Wallenberg Foundation.

This work is based on data products from observations made with ESO Telescopes at the La Silla Paranal Observatory under programme ID 188.B-3002. These data products have been processed by the Cambridge Astronomy Survey Unit (CASU) at the Institute of Astronomy, University of Cambridge, and by the FLAMES/UVES reduction team at INAF/Osservatorio Astrofisico di Arcetri. These data have been obtained from the {\it Gaia}-ESO Survey Data Archive, prepared and hosted by the Wide Field Astronomy Unit, Institute for Astronomy, University of Edinburgh, which is funded by the UK Science and Technology Facilities Council.

This work was partly supported by the European Union FP7 programme through ERC grant number 320360 and by the Leverhulme Trust through grant RPG-2012-541. We acknowledge the support from INAF and Ministero dell' Istruzione, dell' Universit\`a' e della Ricerca (MIUR) in the form of the grant "Premiale VLT 2012". The results presented here benefit from discussions held during the {\it Gaia}-ESO workshops and conferences supported by the ESF (European Science Foundation) through the GREAT Research Network Programme.

The authors thank the anonymous referee for the useful comments that improved the manuscript.

\begin{appendix}

\section{Intra-Node chemical species corrections}\label{appsec:abuncor}

\begin{figure*}[h!]
\centering
\includegraphics[width=18.0cm]{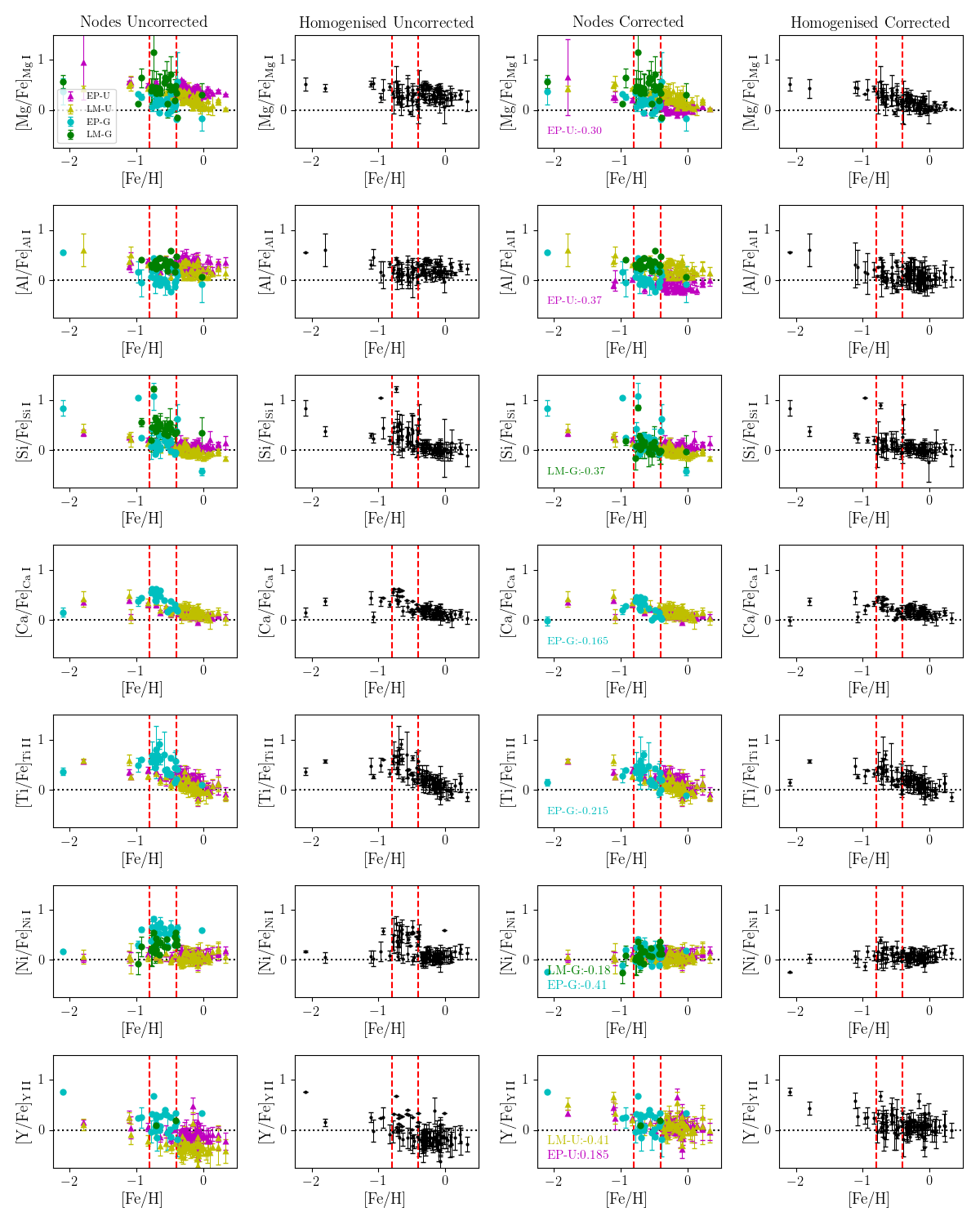}
\caption{Chemical abundances of [X/Fe] against \feh{} for Mg\,\textsc{i}, Al\,\textsc{i}, Si\,\textsc{i}, Ca\,\textsc{i}, Ti\,\textsc{ii}, Ni\,\textsc{i,} and Y\,\textsc{ii} for the K2@{\it Gaia}-ESO stars derived based on the final stellar parameters. The first column shows the node (EP,LM) and instrument (U,G) results as specified in the legend. The \feh{} overlap is bounded by dashed red lines. The errorbars are the line-by-line values per star per node uncertainties. The second column shows the homogenised abundances without per node corrections. The third column shows the node values with corrections applied. The fourth column show the final homogenised values.}
\label{fig:corabund}
\end{figure*}

\section{K2@Gaia-ESO catalogue columns}\label{appsec:fincols}
\begin{table}[htbp]
\scriptsize{
  \setlength\tabcolsep{3.0pt} 
  \centering
  \caption{Columns contained in the K2@Gaia-ESO Catalogue. Full table is available online.}
    \begin{tabular}{ll}
    \hline\hline
    Column name & Description \\
    \hline
    CNAME & Gaia-ESO object name from coordinates \\
    EPIC\_ID & K2 unique identifier \\
    INSTRUMENT & Instrument used for spectroscopic observation \\
    SNR & Signal-to-noise ratio of Gaia-ESO spectrum \\
    GAIADR2\_ID & Gaia DR2 Source Identifier \\
    GAIADR2\_MAGG & Gaia DR2 G band magnitude [Vega] \\
    GAIADR2\_PARAL & Gaia DR2 Parallax \\
    GAIADR2\_PARAL\_ERR & Uncertainty on GAIADR2\_PARAL \\
    TEFF  & Spectroscopic effective temperature \\
    TEFF\_ERR & Error on TEFF \\
    LOGG  & Seismic surface gravity \\
    LOGG\_ERR & Error on LOGG \\
    FEH   & Spectroscopic metallicity \\
    FEH\_ERR & Error on FEH \\
    XI    & Spectroscopic microturbulence \\
    XI\_ERR & Uncertainty on XI \\
    NODE\_RES & Node results used for final parameters \\
    GAIADR2\_VRAD & Gaia DR2 radial velocity \\
    GAIADR2\_VRAD\_ERR & Uncertainty on GAIADR2\_VRAD \\
    VRAD  & Radial Velocity from Gaia-ESO \\
    VRAD\_ERR & Uncertainty on VRAD \\
    VSINI & Rotational Velocity from Gaia-ESO \\
    BIN\_FLAG & Binary detected? 0=no; 1=potential, -1=NA \\
    NA1   & Neutral Sodium Abundance [Na/Fe] \\
    NA1\_ERR & Uncertainty on NA1 \\
    MG1   & Neutral Magnesium Abundance [Mg/Fe] \\
    MG1\_ERR & Uncertainty on MG1 \\
    AL1   & Neutral Aluminium Abundance [Al/Fe] \\
    AL1\_ERR & Uncertainty on AL1 \\
    SI1   & Neutral Silicon Abundance [Si/Fe] \\
    SI1\_ERR & Uncertainty on SI1 \\
    SI2   & Ionised Silicon Abundance [Si/Fe] \\
    SI2\_ERR & Uncertainty on SI2 \\
    CA1   & Neutral Calcium Abundance [Ca/Fe] \\
    CA1\_ERR & Uncertainty on CA1 \\
    CA2   & Ionised Calcium Abundance [Ca/Fe] \\
    CA2\_ERR & Uncertainty on CA2 \\
    SC1   & Neutral Scandium Abundance [Sc/Fe] \\
    SC1\_ERR & Uncertainty on SC1 \\
    SC2   & Ionised Scandium Abundance [Sc/Fe] \\
    SC2\_ERR & Uncertainty on SC2 \\
    TI1   & Neutral Titanium Abundance [Ti/Fe] \\
    TI1\_ERR & Uncertainty on TI1 \\
    TI2   & Ionised Titanium Abundance [Ti/Fe] \\
    TI2\_ERR & Uncertainty on TI2 \\
    V1    & Neutral Vanadium Abundance [V/Fe] \\
    V1\_ERR & Uncertainty on V1 \\
    CR1   & Neutral Chromium Abundance [Cr/Fe] \\
    CR1\_ERR & Uncertainty on CR1 \\
    CR2   & Ionised Chromium Abundance [Cr/Fe] \\
    CR2\_ERR & Uncertainty on CR2 \\
    MN1   & Neutral Manganese Abundance [Mn/Fe] \\
    MN1\_ERR & Uncertainty on MN1 \\
    FE1   & Neutral Iron Abundance [Fe/H] \\
    FE1\_ERR & Uncertainty on FE1 \\
    FE2   & Ionised Iron Abundance [Fe/H] \\
    FE2\_ERR & Uncertainty on FE2 \\
    CO1   & Neutral Cobalt Abundance [Co/Fe] \\
    CO1\_ERR & Uncertainty on CO1 \\
    NI1   & Neutral Nickel Abundance [Ni/Fe] \\
    NI1\_ERR & Uncertainty on NI1 \\
    CU1   & Neutral Copper Abundance [Cu/Fe] \\
    CU1\_ERR & Uncertainty on CU1 \\
    ZN1   & Neutral Zinc Abundance [Zn/Fe] \\
    ZN1\_ERR & Uncertainty on ZN1 \\
    Y2    & Ionised Yttrium Abundance [Y/Fe] \\
    Y2\_ERR & Uncertainty on Y2 \\
    ZR1   & Neutral Zirconium Abundance [Zr/Fe] \\
    ZR1\_ERR & Uncertainty on ZR1 \\
    ZR2   & Ionised Zirconium Abundance [Zr/Fe] \\
    ZR2\_ERR & Uncertainty on ZR2 \\
    BA2   & Ionised Barium Abundance [Ba/Fe] \\
    BA2\_ERR & Uncertainty on BA2 \\
    LA2   & Ionised Lanthanum Abundance [La/Fe] \\
    LA2\_ERR & Uncertainty on LA2 \\
    CE2   & Ionised Cerium Abundance [Ce/Fe] \\
    CE2\_ERR & Uncertainty on CE2 \\
    ND2   & Ionised Neodymium Abundance [Nd/Fe] \\
    ND2\_ERR & Uncertainty on ND2 \\
    EU2   & Ionised Europium Abundance [Eu/Fe] \\
    EU2\_ERR & Uncertainty on EU2 \\
    MASS  & Seismic stellar mass \\
    MASS\_ERR & Error on MASS \\
    AGE   & Seismic stellar age \\
    AGE\_ERR & Error on AGE \\
    \hline
    \end{tabular}%
  \label{tab:onelinetab_cols}%
  }
\end{table}%

\begin{table}[htbp]
\scriptsize{
  \setlength\tabcolsep{3.0pt} 
  \centering
  \caption{Columns for line-by-line per node per cname abundance analysis. Full table is available online.}
    \begin{tabular}{ll}
    \hline\hline
    Column name & Description \\
    \hline
    NODE  & Gaia-ESO node name \\
    CNAME & Gaia-ESO object name from coordinates \\
    INSTRUMENT & Spectroscopic instrument \\
    LAMBDA & Wavelength of spectral line \\
    ELEMENT & Atomic element name \\
    ION   & Species: 1=neutral, 2=ionised \\
    EXC\_POT & Excitation potential \\
    LOG\_GF & Oscillator strength \\
    REFERENCE & Atomic information reference \\
    EW    & Measured equivalent width \\
    ABUND & Measured abundance as log(eps) \\
    ABUND\_ERR & Error on ABUND \\
    ABUND\_UPPER & Limit flag: 0=Detection; 1=Upper limit \\
    MEAS\_TYPE & SS: Spectrum Synthesis; EW: Equivalent Widths \\
    \hline
    \end{tabular}%
  \label{tab:lbltab_cols}%
 }
\end{table}%

\section{K2@Gaia-ESO chemical abundance distributions}\label{appsec:abund}
\begin{figure*}
\centering
\includegraphics[width=17cm]{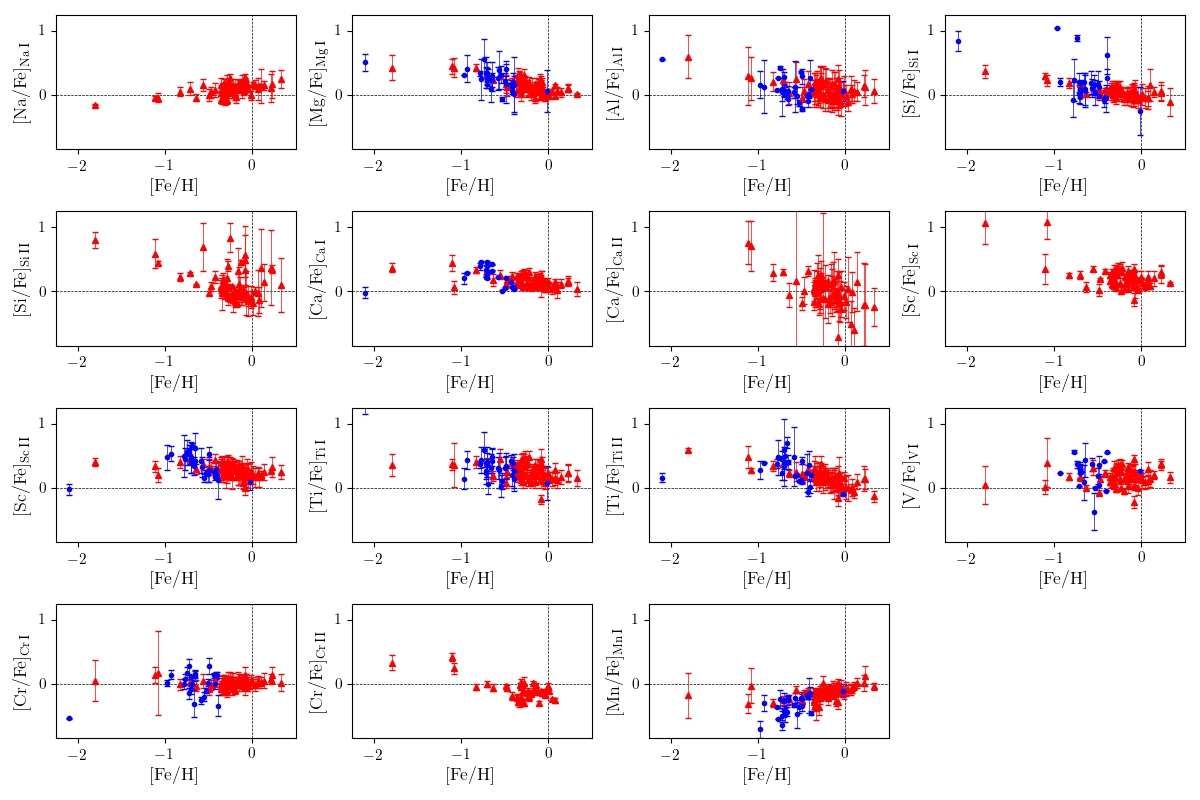}
\includegraphics[width=17cm]{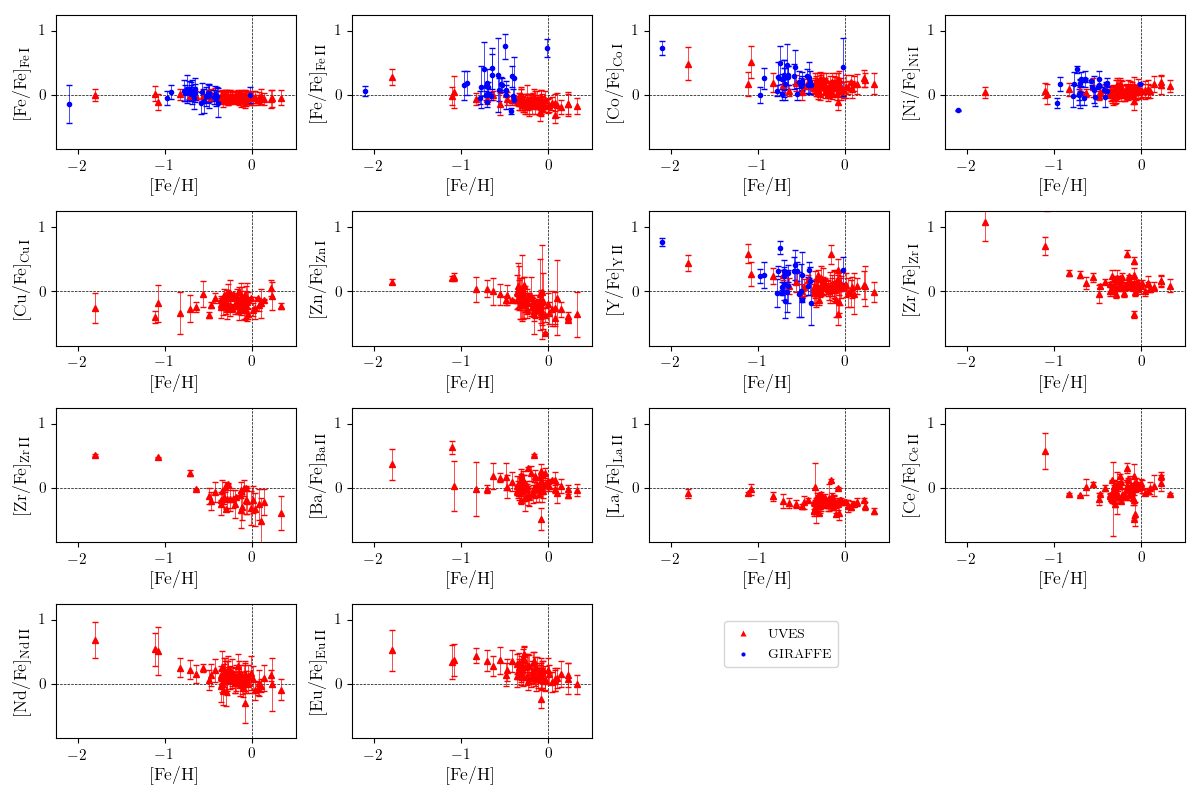}
\caption{Chemical abundances of [X/Fe] against \feh{} for the K2@{\it Gaia}-ESO stars derived based on the final stellar parameters. The UVES stars are red triangles, the GIRAFFE stars are blue circles. \feh{}=0 and [X/Fe]=0 are indicated as dashed lines.}
\label{fig:finabund}
\end{figure*}

\section{ {\it Gaia} surface gravity}\label{app:gaialogg}
Bolometric corrections were interpolated from the tables of \cite{Houdashelt2000}. These corrections are tabulated for the K magnitude in the CIT/CTIO system \citep{Elias1982}. Thus, first the Ks magnitude of 2MASS were converted to K$_{\textrm{CIT/CTIO}}$ using the relations provided by \cite{Carpenter2001}.

The bolometric corrections (BC) are mostly independent of \logg{}. First the two tables of metallicity values closest to the metallicity of the star (usually \feh{} = 0.00 and \feh{} = -0.50) are found. These tables are used to linearly interpolate the corrections to the \teff{} of the star. Then a second linear interpolation is carried out to the metallicity of the star.

The E(B-V) is converted to A$_{\textrm{Ks}}$ using the coefficients of \cite{McCall2004}. In this case, A$_{\textrm{Ks}}$/E(B-V) = 0.350.

The absolute magnitude (Mabs) is then:
\begin{equation}
    \textrm{M}abs = (\textrm{K}_{\textrm{2MASS}} - A_{\textrm{Ks}}) + 0.024 + 5 + 5 \log_{10} (0.001 \pi)
,\end{equation}

where +0.024 is the conversion to K$_{\textrm{CIT/CTIO}}$ and the parallax ($\pi$) has to be in arcsec (hence the 0.001)

This implicitly assumes that distance = 1/parallax which is only then calculated if the relative uncertainty of the parallax is better than 15\%.

Then the bolometric magnitude (Mbol) and luminosity of the star relative to the Sun $\log(\textrm{L}_{*}/\textrm{L}_\odot)$ are:
\begin{eqnarray}
\textrm{M}bol & = & \textrm{M}abs + \textrm{BC}  \\
\log(\textrm{L}_{*}/\textrm{L}_\odot) & = & -0.4~(\textrm{M}bol-4.75)
\end{eqnarray}

where 4.75 is the bolometric magnitude of the Sun.

And finally:

{\footnotesize
\begin{equation}
    \log g = 4.44 + 4 \log_{10}(T_{\mathrm{eff}}/5771) - \log_{10}(\textrm{L}_{*}/\textrm{L}_{\odot}) + \log_{10}(\textrm{M}_{*}/\textrm{M}_\odot)
\end{equation}\label{eqn:logg}
}

The stellar masses were then estimated using the {\it UniDAM} code  \citep[see http://www2.mps.mpg.de/homes/mints/unidam.html][]{Mints2017}. It interpolates masses from PARSEC isochrones \citep{PARSEC} using a Bayesian scheme. A first estimate of mass was computed with this code using the spectroscopic parameters (\teff{}, \logg{}, \feh{}), 2MASS magnitudes, parallaxes, and extinction.

With this first mass estimate, we compute a first estimate of \logg{} using Equation~\ref{eqn:logg}. The new \logg{} values are then given again to UniDAM, to recompute the mass estimates. These iterations were repeated (a total of four iterations was sufficient) until the estimated mass was consistent with the estimated \logg{}.

UniDAM gives an estimate of the error of the mass, and this was also used in the computation of the total error budget in \logg{}. Uncertainties are from 10000 Monte Carlo simulations, assuming Gaussian uncertainties in $\pi$, K$_{\textrm{2MASS}}$, stellar mass (with error coming from UniDAM), and $\pm 0.02$~mag for the bolometric correction. No uncertainty was assumed in E(B-V), nor for \teff{}$_\odot$, or \logg{}$_\odot$.

\end{appendix}

\clearpage
\onecolumn

\setcounter{table}{0}

{
\footnotesize
\begin{longtable}{llcc|ccc|c|cc|c}
\caption{The table of the 90 stars analysed in the K2 {\it Gaia}-ESO special project. Listed are: the {\it Gaia}-ESO CNAME; EPIC identifier; the spectroscopic instrument used; the IRFM \teff{}; the preliminary set of spectroscopic parameters (Prelim. Spec.) sourced from the iDR5 recommended parameters where available (noted as *) or otherwise parameters from the Nice Node analysis or the GIRAFFE reduction pipeline radial velocity determination; the preliminary seismic \logg{}; the initial spectroscopic \teff{} and \feh{} from iSpec; the initial seismic \logg{}.}  \\
\hline\hline
          &       &       & \multicolumn{1}{c}{IRFM} & \multicolumn{3}{|c}{Prelim. Spec.} & \multicolumn{1}{|c}{Prelim. Seis.} & \multicolumn{2}{|c}{Initial Spec. (iSpec)} & \multicolumn{1}{|c}{Initial Seis.} \\

    \multicolumn{1}{c}{CNAME} & \multicolumn{1}{c}{EPIC ID} & \multicolumn{1}{c}{ Instrument} & \multicolumn{1}{c}{\teff{}} & \multicolumn{1}{|c}{\teff{}} & \multicolumn{1}{c}{\logg{}} & \multicolumn{1}{c}{\feh{}} & \multicolumn{1}{|c}{\logg{}} & \multicolumn{1}{|c}{\teff{}} & \multicolumn{1}{c}{\feh{}} & \multicolumn{1}{|c}{\logg{}} \\

     &  &  & [K] & [K] & [$\log(\textrm{cms}^{-2})$] &  & [$\log(\textrm{cms}^{-2})$] & [K] &  & [$\log(\textrm{cms}^{-2})$] \\

\hline 
\endfirsthead
\caption{continued.}\\
\hline\hline
          &       &       & \multicolumn{1}{c}{IRFM} & \multicolumn{3}{|c}{Prelim. Spec.} & \multicolumn{1}{|c}{Prelim. Seis.} & \multicolumn{2}{|c}{Initial Spec. (iSpec)} & \multicolumn{1}{|c}{Initial Seis.} \\

    \multicolumn{1}{c}{CNAME} & \multicolumn{1}{c}{EPIC ID} & \multicolumn{1}{c}{ Instrument} & \multicolumn{1}{c}{\teff{}} & \multicolumn{1}{|c}{\teff{}} & \multicolumn{1}{c}{\logg{}} & \multicolumn{1}{c}{\feh{}} & \multicolumn{1}{|c}{\logg{}} & \multicolumn{1}{|c}{\teff{}} & \multicolumn{1}{c}{\feh{}} & \multicolumn{1}{|c}{\logg{}} \\

     &  &  & [K] & [K] & [$\log(\textrm{cms}^{-2})$] &  & [$\log(\textrm{cms}^{-2})$] & [K] &  & [$\log(\textrm{cms}^{-2})$] \\

\hline 
\endhead
\hline
\endfoot
        22013945-0909229  &  K2\_206238956  &  UVES*  & 4541  & 4607  & 3.15  & 0.18  & 2.98  & 4721  & 0.25  & 3.04 \\
        22032202-0829154  &  K2\_206283732  &  UVES*  & 5125  & 5089  & 3.84  & 0.23  & 2.45  & 5421  & 0.29  & 2.49 \\
        22050764-0840472  &  K2\_206270701  &  UVES*  & 4892  & 4935  & 3.31  & -0.26 & 3.07  & 4992  & -0.24 & 3.11 \\
        22052114-0848002  &  K2\_206262569  &  UVES*  & 4697  & 4762  & 2.66  & -0.32 & 2.41  & 4820  & -0.32 & 2.46 \\
        22052550-0854435  &  K2\_206255139  &  UVES*  & 4756  & 4776  & 2.43  & -0.29 & 2.31  & 4876  & -0.28 & 2.36 \\
        22055148-0840028  &  K2\_206271510  &  UVES*  & 4593  & 4689  & 3.00  & -0.29 & 2.63  & 4800  & -0.21 & 2.68 \\
        22065112-1504580  &  K2\_205991117  &  UVES*  & 4798  & 4899  & 1.96  & -1.68 & 1.99  & 4602  & -1.89 & 2.01 \\
        22071748-1455524  &  K2\_205996255  &  UVES*  & 4641  & 4707  & 2.74  & -0.26 & 2.51  & 4795  & -0.22 & 2.56 \\
\end{longtable}
}

{
\scriptsize
\setlength\tabcolsep{2.5pt} 
\begin{landscape}
\begin{longtable}{lc|ccccccc|ccccccc|cc|ccccc}
    \caption{Parameters determined by IRFM, the final parameters from each node, the outlier assessment from uncertainties, and the final parameters for the 62 UVES and 28 GIRAFFE stars.}\label{tab:uves-gir_outliers}\\
    
\hline\hline
    & IRFM  & \multicolumn{7}{|c}{EPINARBO (EP)}        & \multicolumn{7}{|c}{Lumba (LM)}  & \multicolumn{2}{|c}{$\Delta T_{\mathrm{eff}}$} & \multicolumn{5}{|c}{LM-EP} \\
    CNAME & $T_{\mathrm{eff}}$ & $\log g$ & $T_{\mathrm{eff}}$ & $\sigma$ & [Fe/H] & $\sigma$ & $\xi$ & $\chi^2$ & $\log g$ & $T_{\mathrm{eff}}$ & $\sigma$ & [Fe/H] & $\sigma$ & $\xi$ & $\chi^2$ & IRFM-EP & IRFM-LM & $\Delta\log g$ & $\Delta T_{\mathrm{eff}}$ & $\Delta$[Fe/H] & $\Delta\xi$ & $\Delta\chi^2$ \\
    
       &  [K] & 
    [{\tiny $\log(\mathrm{cms}^{-2})$}] & [K]  & [K] &  &  & [km/s] & [{\tiny $\log(\mathrm{cms}^{-2})$}] &
    [{\tiny $\log(\mathrm{cms}^{-2})$}] & [K]  & [K] &  &  & [km/s] & [{\tiny $\log(\mathrm{cms}^{-2})$}] &
    [K]   &  [K] & 
    [{\tiny $\log(\mathrm{cms}^{-2})$}] & [K] &  & [km/s] & [{\tiny $\log(\mathrm{cms}^{-2})$}] \\
    
\hline
\endfirsthead

\caption{continued.}\\
\hline\hline
    & IRFM  & \multicolumn{7}{|c}{EPINARBO (EP)}        & \multicolumn{7}{|c}{Lumba (LM)}  & \multicolumn{2}{|c}{$\Delta T_{\mathrm{eff}}$} & \multicolumn{5}{|c}{LM-EP} \\
    CNAME & $T_{\mathrm{eff}}$ & $\log g$ & $T_{\mathrm{eff}}$ & $\sigma$ & [Fe/H] & $\sigma$ & $\xi$ & $\chi^2$ & $\log g$ & $T_{\mathrm{eff}}$ & $\sigma$ & [Fe/H] & $\sigma$ & $\xi$ & $\chi^2$ & IRFM-EP & IRFM-LM & $\Delta\log g$ & $\Delta T_{\mathrm{eff}}$ & $\Delta$[Fe/H] & $\Delta\xi$ & $\Delta\chi^2$ \\
    
       &  [K] & 
    [{\tiny log(cms$^{-2}$)}] & [K]  & [K] &  &  & [km/s] & [{\tiny $\log(\mathrm{cms}^{-2})$}] &
    [\tiny{$\log(\textrm{cms}^{-2})$}] & [K]  & [K] &  &  & [km/s] & [\tiny{$\log(\textrm{cms}^{-2})$}] &
    [K]   &  [K] & 
    [\tiny{$\log(\textrm{cms}^{-2})$}] & [K] &  & [km/s] & [\tiny{$\log(\textrm{cms}^{-2})$}] \\

\hline 
\endhead
\hline
\endfoot
     &   &   &   &   &  &   &   &       &    &    &     &   &    &    &    &  &    &    &     &   &   &  \\
    \multicolumn{1}{l}{\bf UVES} &   &   &   &   &  &   &   &       &    &    &     &   &    &    &    &  &    &    &     &   &   &  \\
    22172723-1633039 & 4482 & 1.95 & 4653 & 150 & -0.76 & 0.09 & 1.44 & ...     & 1.95  & 4631  & 168   & -0.89 & 0.21  & 1.32  & ...     & -171  & -149  & 0.00  & -22   & -0.13 & -0.12 & ... \\
    22075605-1535081 & 4669 & 2.85 & 4882 & 150 & 0.05 & 0.11 & 1.36 & ...     & 2.85  & 4762  & 116   & -0.12 & 0.16  & 1.24  & ...     & -213  & -93   & -0.01 & -120  & -0.17 & -0.12 & ... \\
    22072959-1530438 & 4884 & 3.05 & 5051 & 150 & -0.05 & 0.10 & 1.26 & ...     & 3.04  & 4974  & 166   & -0.22 & 0.21  & 1.16  & ...     & -167  & -90   & 0.00  & -77   & -0.17 & -0.10 & ... \\
    22373392-1521214 & 4736 & 2.80 & 4924 & 150 & -0.03 & 0.11 & 1.41 & ...     & 2.80  & 4827  & 131   & -0.18 & 0.16  & 1.21  & ...     & -188  & -91   & 0.00  & -97   & -0.15 & -0.20 & ... \\
    22370810-1514265 & 4458 & 2.49 & 4697 & 150 & -0.22 & 0.11 & 1.44 & ...     & 2.48  & 4623  & 137   & -0.32 & 0.19  & 1.32  & ...     & -239  & -165  & 0.00  & -74   & -0.10 & -0.12 & ... \\
    22065112-1504580 & 4798 & 2.01 & 4604 & 150 & -1.89 & 0.11 & 1.36 & ...     & 2.02  & 4875  & 233   & -1.70 & 0.26  & 1.20  & ...     & 194   & -77   & 0.01  & 271   & 0.19  & -0.16 & ... \\
\end{longtable}
\end{landscape}
}

{
\footnotesize
\setlength\tabcolsep{3.0pt} 
\begin{longtable}{llcccccccccc}
    \caption{Final stellar parameters for the 90 K2@{\it Gaia}-ESO stars.}\label{tab:fin_stellar_params}\\
\hline\hline
    CNAME & EPIC ID & Instrument & Node SP & \multicolumn{1}{c}{$T_{\mathrm{eff}}$} & \multicolumn{1}{c}{$\sigma$} & \multicolumn{1}{c}{$\log g$} & \multicolumn{1}{c}{$\sigma$} & \multicolumn{1}{c}{[Fe/H]} & \multicolumn{1}{c}{$\sigma$} & \multicolumn{1}{c}{$\xi$} & \multicolumn{1}{c}{$\sigma$} \\

     &   &   &   & [K] & [K] & [{\tiny $\log(\mathrm{cms}^{-2})$}] & [{\tiny $\log(\mathrm{cms}^{-2})$}] &  &  & [km/s] & [km/s] \\

\hline
\endfirsthead

\caption{continued.}\\
\hline\hline
    CNAME & EPIC ID & Instrument & Node SP & \multicolumn{1}{l}{$T_{\mathrm{eff}}$} & \multicolumn{1}{l}{$\sigma$} & \multicolumn{1}{l}{$\log g$} & \multicolumn{1}{l}{$\sigma$} & \multicolumn{1}{l}{[Fe/H]} & \multicolumn{1}{l}{$\sigma$} & \multicolumn{1}{l}{$\xi$} & \multicolumn{1}{l}{$\sigma$} \\
    \\
\hline 
\endhead
\hline
\endfoot
    21563608-1202424 & K2\_206102116 & UVES  & Both  & 4850  & 11    & 3.01  & 0.02  & -0.28 & 0.10   & 1.18  & 0.02 \\
    21590887-1159078 & K2\_206104478 & UVES  & Both  & 4741  & 93    & 2.45  & 0.02  & -0.07 & 0.11  & 1.40   & 0.11 \\
    22000793-1203412 & K2\_206101493 & UVES  & EP    & 5309  & 150   & 2.46  & 0.02  & -1.08 & 0.14  & 2.00   & 0.15 \\
    22013369-1141245 & K2\_206115828 & UVES  & Both  & 4688  & 62    & 2.33  & 0.03  & -0.47 & 0.13  & 1.37   & 0.06 \\
    22013945-0909229 & K2\_206238956 & UVES  & Both  & 4695  & 37    & 3.13  & 0.02  & 0.22  & 0.01  & 1.41   & 0.12 \\
    22015504-1153022 & K2\_206108556 & UVES  & Both  & 4802  & 66    & 2.57  & 0.02  & -0.28 & 0.10   & 1.35  & 0.10 \\
    22021848-1139147 & K2\_206117168 & UVES  & Both  & 4700  & 132   & 2.48  & 0.02  & 0.07  & 0.06  & 1.49  & 0.16 \\
    22032202-0829154 & K2\_206283732 & UVES  & LM    & 4742  & 159   & 2.55  & 0.02  & -0.08 & 0.18  & 1.25  & 0.07 \\
\end{longtable}
}

\setcounter{table}{5}

{
\footnotesize
\setlength\tabcolsep{3.0pt} 
\begin{longtable}{ll|rcc|rcc|c}
\caption{Radial (RV) and rotational velocities ($v \sin i$) of the K2@{\it Gaia}-ESO stars measured by {\it Gaia}-ESO and from the {\it Gaia} Data Release 2 (GDR2). NT refers to the number of transits {\it Gaia} considered for the RV measurement. A BIN\_FLAG of 1 indicates a possible binary based on the GDR2 and {\it Gaia}-ESO measurements (see Sect.~\ref{sect:binaries}).}\label{6xGaia DR2}\\
\hline\hline
  \multicolumn{1}{c}{CNAME} &
  \multicolumn{1}{c|}{GDR2\_ID} &
  \multicolumn{3}{c|}{{\it Gaia}-ESO} &
  \multicolumn{3}{c|}{GDR2} &
  \multicolumn{1}{c}{BIN\_FLAG} \\
   &
   &
  \multicolumn{1}{c}{RV} &
  \multicolumn{1}{c}{$\sigma_{RV}$} &
  \multicolumn{1}{c|}{$v \sin i$} &
  \multicolumn{1}{c}{RV} &
  \multicolumn{1}{c}{$\sigma_{RV}$} &
  \multicolumn{1}{c|}{NT} &
  \multicolumn{1}{c}{} \\
   &
   &
  [km/s] &
  [km/s]  &
  [km/s]  &
  [km/s]  &
  [km/s]  &
   &
   \\
  
\hline
\endfirsthead

\caption{continued.}\\
\hline\hline
  \multicolumn{1}{c}{CNAME} &
  \multicolumn{1}{c|}{GDR2\_ID} &
  \multicolumn{3}{c|}{{\it Gaia}-ESO} &
  \multicolumn{3}{c|}{GDR2} &
  \multicolumn{1}{c}{BIN\_FLAG} \\
   &
   &
  \multicolumn{1}{c}{RV} &
  \multicolumn{1}{c}{$\sigma_{RV}$} &
  \multicolumn{1}{c|}{$v \sin i$} &
  \multicolumn{1}{c}{RV} &
  \multicolumn{1}{c}{$\sigma_{RV}$} &
  \multicolumn{1}{c|}{NT} &
  \multicolumn{1}{c}{} \\
   &
   &
  [km/s] &
  [km/s]  &
  [km/s]  &
  [km/s]  &
  [km/s]  &
   &
   \\
\hline 
\endhead
\hline
\endfoot
    21563608-1202424 &  2613587930052748544  & 21.66 & 0.13  & 3.48   &  ...   &  ...  & 0      & ... \\
      21585236-1201297  &  2613577385908074240  & -67.49 & 0.17  & 3.00   &  ...   &  ...  & 0      & ... \\
      21590887-1159078  &  2613600956688590592  & 16.67 & 0.13  & 3.46  & 16.38 & 0.54  & 7     & 0 \\
      22000793-1203412  &  2613550340498556544  & -0.03 & 0.13  & 2.18  & 1.90  & 1.42  & 6     & 1 \\
      22003290-0808595  &  2618490072580174720  & -2.53 & 0.15  & 3.35  & -4.84 & 2.95  & 5     & 1 \\
      22013286-1140363  &  2613657547177899648  & -139.72 & 0.15  & 2.80   &  ...   &  ...  & 0      & ... \\
      22013369-1141245  &  2613657439802765952  & 19.69 & 0.13  & 2.00  & 20.04 & 0.22  & 3     & 0 \\
      22013945-0909229  &  2617607955016992512  & -33.49 & 0.13  & 4.32  & -33.32 & 0.34  & 6     & 0 \\
\end{longtable}
}

\end{document}